\documentclass[apj]{emulateapj}

\usepackage[dvips]{color}

\slugcomment{Draft version \today}

\begin{document}

%\begin{CJK*}{JIS}{}

\title{Magnetic Field Amplification Associated with the
  Richtmyer-Meshkov Instability}

%\author{Takayoshi Sano}
%\affil{Institute of Laser Engineering, Osaka University, Suita, Osaka
%  565-0871, Japan}
%\email{sano@ile.osaka-u.ac.jp}
%\author{Tsuyoshi Inoue}
%\affil{Department of Physics and Mathematics, Aoyama Gakuin
%  University, Sagamihara, Kanagawa 252-5258, Japan}
%\author{Chihiro Matsuoka}
%\affil{Department of Physics, Ehime University, Matsuyama, Ehime
%  790-8577, Japan}
%\and
%\author{Katsunobu Nishihara}
%\affil{Institute of Laser Engineering, Osaka University, Suita, Osaka
%  565-0871, Japan}

\author{Takayoshi Sano\altaffilmark{1}, 
Katsunobu Nishihara\altaffilmark{1},
Chihiro Matsuoka\altaffilmark{2}, and
Tsuyoshi Inoue\altaffilmark{3}}

%\author{Takayoshi Sano (º´Ì¹¥)\altaffilmark{1}, 
%Katsunobu Nishihara (À¾¸¶¸ù½¤)\altaffilmark{1},
%Chihiro Matsuoka (¾¾²¬ÀéÇî)\altaffilmark{2}, and
%Tsuyoshi Inoue (°æ¾å¹ä»Ö)\altaffilmark{3}}

\altaffiltext{1}{Institute of Laser Engineering, Osaka University,
  Suita, Osaka 565-0871, Japan}
\email{sano@ile.osaka-u.ac.jp}
\altaffiltext{2}{Department of Physics, Ehime University, Matsuyama,
  Ehime 790-8577, Japan}
\altaffiltext{3}{Department of Physics and Mathematics, Aoyama Gakuin
  University, Sagamihara, Kanagawa 252-5258, Japan}

%\end{CJK*}

\begin{abstract}
The amplification of a magnetic field due to the Richtmyer-Meshkov
instability (RMI) is investigated by two-dimensional MHD simulations. 
Single-mode analysis is adopted to reveal definite relation between
the nonlinear evolution of RMI and the field enhancement.   
It is found that an ambient magnetic field is stretched by fluid
motions associated with the RMI, and the strength is amplified
significantly by more than two orders of magnitude. 
The saturation level of the field is determined by a balance between
the amplified magnetic pressure and the thermal pressure after shock
passage.    
This effective amplification can be achieved in a wide range of the
conditions for the RMI such as the Mach number of an incident shock
and the density ratio at a contact discontinuity.   
The results suggest that the RMI could be a robust mechanism of the
amplification of interstellar magnetic fields and cause the origin of
localized strong fields observed at the shock of supernova remnants. 
\end{abstract}


\section{Introduction}

The Richtmyer-Meshkov instability (RMI) is of crucial importance in a
variety of applications including astrophysical phenomena and
laboratory experiments \citep{brouillette02,nishihara10}. 
The RMI occurs when an incident shock strikes a corrugated contact
discontinuity separating two fluids with different densities
\citep{richtmyer60,meshkov69}. 
Because of the corrugation of the interface, the surface profiles of
the transmitted and reflected shocks are also rippled. 
The RMI is driven by the vorticity left by these rippled shocks at the
interface and in the fluids \citep{wouchuk97}.
%The basic mechanism for the RMI is baroclinic vorticity generation at
%the interface resulting from the misalignment of the pressure gradient
%of the shock and the local density gradient across the interface.  

Supernova remnants (SNRs) are expected as a site of magnetic field
amplification. 
Thin shell structure of the front shock in young SNRs observed by hard
X-rays \citep[e.g.,][]{long03,bamba03} requires the existence of
strong fields with $\sim 100$ $\mu$G, which is estimated from the
syncrotron cooling length \citep{berezhko03,volk05}.
Recent discovery of the year-scale variability in the synchrotron
X-ray emission at the downstream regions of supernova shocks
suggests that the magnetic field should be amplified up to the level
of milligauss there \citep{uchiyama07,uchiyama08}. 

In this paper, we consider the origin of this localized milligauss
field in SNRs.
Since the typical magnetic field in the interstellar medium (ISM) is
of the order of microgauss \citep{beck01,heiles05}, the amplification
beyond simple shock compression is necessary to achieve the
milligauss-order magnetic field.   
It is reported that observed synchrotron images could be modeled by
random magnetic fields \citep{bykov08}, but the amplification mechanism
of the field is still an open question.
As discussed in \citet{uchiyama07}, the presence of milligauss
magnetic field has a critical meaning in the long-standing paradigm of
cosmic-ray proton acceleration in young SNRs
\citep{bell78,blandford78}.    
The investigation on the physical origin of such strong field is then
an inevitable step to clarify the role of the milligauss field.

%It is widely known that the interstellar medium (ISM) has nonuniform
%density structures from hot ionized medium \citep{armstrong95} to warm
%and cold neutral media formed by the thermal instability
%\citep{field69,koyama02,inoue09}.    

It is widely known that the ISM has nonuniform density structures 
of warm and cold neutral media formed by the thermal
instability \citep{field69,koyama02,inoue09}.  
The electron density fluctuation in hot ionized medium is also
observed as an evidence of the interstellar turbulence
\citep{armstrong95}. 
The RMI can take place when supernova shocks pass through these
inhomogeneous materials, and which is a quite common event in the ISM. 
The low- and mid-temperature parts of the ISM is partially ionized by
cosmic rays.  
Although the ionization fraction is very small, it is well described
by the ideal MHD approximation.  
%Hence the fluid motions of RMI in the ISM could be affected by a
%magnetic field.  
Hence the interstellar magnetic fields could be affected by the fluid
motions of the RMI. 

Turbulence and magnetic field amplification has been studied for shock
wave propagation through the inhomogeneous ISM
\citep{giacalone07,inoue09,inoue12,guo12}.
%two-phase ISM composed of small-scale cloudlets \citep{inoue09}.  
%One of the most promising processes of the field amplification is the
%RMI.  
%The amplification of the field related to supernova shocks is also
%pointed out to explain the strong field generation in the
%core-collapse phase \citep{endeve10} and the interaction with the 
%termination shock of the progenitor wind \citep{chevalier92a}. 
Field amplification is inferred to take place both in the star's
collapsing core \citep{endeve10} and in the progenitor wind overrun by
the shock \citep{chevalier92a}, and in both cases the shock itself has
been suggested as the cause of the amplification. 
%In this paper, we examine whether the RMI can amplify the interstellar
%field by a factor of more than 100 using two-dimensional MHD
%simulations. 
\citet{inoue12} examined the three-dimensional
magnetohydrodynamical interaction between clumpy interstellar clouds 
formed by the thermal instability and a strong shock wave of young
SNRs with the typical velocity $\sim$ 3000 km s$^{-1}$.   
It is found that the maximum strength of the interstellar field
increases rapidly owing to the shear layer at the cloud interface, in
addition to rather gradual growth of the average field strength by the
so-called turbulent dynamo (or turbulent stretching of magnetic
field).   
The maximum field strength reaches 1 mG in their simulations, which is
larger by an order than the volume averaged strength. 
The strong-field regions are localized in thin layers at the
interface.

Turbulent dynamo is one of the promising effects to amplify the
magnetic field in the postshocked regions. 
This process dominates the growth of the volume averaged field
strength, which has been recognized by many authors
\citep{giacalone07,mizuno11}. 
On the other hand, the size of the strong-field regions inferred from
X-ray observations is typically 0.05 pc \citep{uchiyama07}, and this
would be much smaller than the preshock inhomogeneity scale $\sim$ 1
pc.  
Such localized strong field should be attributed to the shock-cloud
interaction, and characterized by the maximum strength rather than the
average.  
Therefore, the evolution of the maximum field strength is a key to
understand the formation of the localized milligauss regions. 
In this paper, we demonstrate the RMI could be an efficient mechanism
of the field amplification at the cloud interface, which can enhance
the maximum strength locally by a factor of more than 100. 

%\citet{inoue12} investigated the magnetohydrodynamical interaction
%between clumpy interstellar clouds formed by the thermal instability
%and a strong shock. 
%It is found that the maximum strength of the interstellar
%field increases rapidly in contrast to rather gradual growth of the
%average field strength. 
%The maximum field is much larger than the average, because the
%strong-field regions are generally localized in tiny thin layers.   
%Turbulent dynamo is one of the promising effects to amplify the
%magnetic field in the postshock turbulence mostly in the warm diffuse
%gas. 
%This can contribute the growth of the average field strength, which
%has been recognized by many authors \citep{giacalone07,mizuno11}.
%%It is suggested that the magnetic field is amplified locally at the
%%interface between dense clouds and diffuse gas. 
%On the other hand, the size of the strong-field regions inferred from
%X-ray observations is typically 0.05 pc, and this would be much
%smaller than the preshock inhomogeneity scale $\sim$ 1 pc. 
%Such localized strong field should be characterized by the maximum strength
%rather than the average. 
%Therefore, the evolution of the maximum field strength is a key 
%to understand the formation of the localized milligauss regions. 
%In this paper, we demonstrate the RMI could be an efficient mechanism
%of the field amplification at the interface, which can enhance the
%maximum strength locally by a factor of more than 100. 

Shock-cloud interaction with a magnetic field has been studied
numerically by multi-dimensional simulations
\citep[e.g.,][]{vanloo07,shin08}.  
In their works, the time evolutions of the entire region of a
round-shaped cloud are considered.  
In contrast, we focus on the hydrodynamical instability arising from
the spacial fluctuations at the cloud surface aiming to see much
smaller scale structures.
The advantage of this approach is that we can make use of the
quantitative knowledge of RMI for the interpretation of the results.

MHD effects on the RMI are investigated firstly in the context of the
dynamics of the magnetosphere \citep{wu99}. 
\citet{samtaney03} has demonstrated by numerical simulations that the
growth of the RMI is suppressed in the presence of a magnetic field. 
Depending on the field direction relative to the incident shock
surface, the magnetic field works differently on the fluid motions of
RMI. 
When the magnetic field is aligned with the motion of the shock,
the suppression of the instability is caused by changes in the shock
refraction process at the contact discontinuity.
The role of the magnetic field is to prevent the deposition of
circulation on the interface \citep{wheatley05,wheatley09}. 
When the magnetic field is parallel to the shock surface, the
instability is stabilized by the Lorentz force \citep{cao08}.

Previous works on the MHD RMI is mostly focused on the suppression
effects by a strong magnetic field.
Here, we concentrate our discussions on the evolution of an ambient
magnetic field affected by the nonlinear motions of the RMI. 
The velocity of the supernova shocks is typically 100-1000 km
s$^{-1}$, and then the Mach number can be 10-100 in the warm diffuse
gas. 
The thermal instability produces the density gap between the cold and
warm neutral media and the ratio is about 10-100. 
Thus, the RMI associated with the supernova shocks is characterized by
a large Mach number and a large density ratio.
In this paper, we show the results of MHD simulations performed with a
variety of initial conditions including such extreme cases.
%In this paper, we perform a variety of models including such extreme
%situations. 

Single-mode analysis of the RMI is adopted in order to examine the
field evolution and understand the physical mechanism.
The plan of this paper is as follows.  
In \S 2, the basic equations, initial conditions, and numerical method
are described.  
In this section, we make a brief review on the characteristics of the
RMI, which is useful for the interpretation of our results. 
Numerical outcomes of two-dimensional MHD simulations are shown in \S
3. 
%In this paper, we focus on the amplification process of a magnetic
%field associated with the growth of the RMI.
The time evolutions of the field for various initial conditions are
examined in this section.
In \S 4, the saturation mechanism of the amplified field and parameter
dependence of the results are discussed.  
%In \S 4, the saturation mechanism and probability distribution
%function of the amplified field are discussed.  
Finally, our conclusions are summarized in \S 5.

\section{Method}

\subsection{Basic Equations and Initial Conditions}

To study the nonlinear evolutions of the RMI, the equations of
ideal MHD are solved;
\begin{equation}
\frac{\partial \rho}{\partial t} + {\mbox{\boldmath{$\nabla$}}} \cdot
\left( \rho {\mbox{\boldmath{$v$}}} \right) = 0 \;,
\end{equation}
\begin{equation}
\frac{\partial \rho {\mbox{\boldmath{$v$}}}}{\partial t} +
{\mbox{\boldmath{$\nabla$}}} \cdot
\left[ \left( P + \frac{B^2}{8 \pi} \right) {\mbox{\boldmath{$I$}}}
+ \rho {\mbox{\boldmath{$v$}}}
{\mbox{\boldmath{$v$}}} - \frac{
{\mbox{\boldmath{$B$}}}{\mbox{\boldmath{$B$}}}}{4 \pi} \right] = 0 \;,
\end{equation}
\begin{equation}
\frac{\partial e}{\partial t} + {\mbox{\boldmath{$\nabla$}}} \cdot
\left[ \left( e + P + \frac{B^2}{8 \pi} \right) 
{\mbox{\boldmath{$v$}}} - \frac{
\left( {\mbox{\boldmath{$B$}}} \cdot {\mbox{\boldmath{$v$}}}
\right) {\mbox{\boldmath{$B$}}}}{4 \pi} 
\right] = 0 \;,
\end{equation}
\begin{equation}
\frac{\partial {\mbox{\boldmath{$B$}}}}{\partial t} = 
{\mbox{\boldmath{$\nabla$}}} \times
\left( {\mbox{\boldmath{$v$}}} \times {\mbox{\boldmath{$B$}}}
\right) \;,
\end{equation}
where $\rho$, {\mbox{\boldmath{$v$}}}, and {\mbox{\boldmath{$B$}}} are
the mass density, velocity, and magnetic field, respectively, and $e$
is the total energy density,
\begin{equation}
e = \frac{P}{\gamma -1} + \frac{\rho v^2}2 + \frac{B^2}{8 \pi} \;.
\end{equation}
In our simulations, the equation of state for ideal gas is adopted
with the adiabatic index $\gamma = 5/3$. 

\begin{figure}
\begin{center}
\includegraphics[scale=0.95,clip]{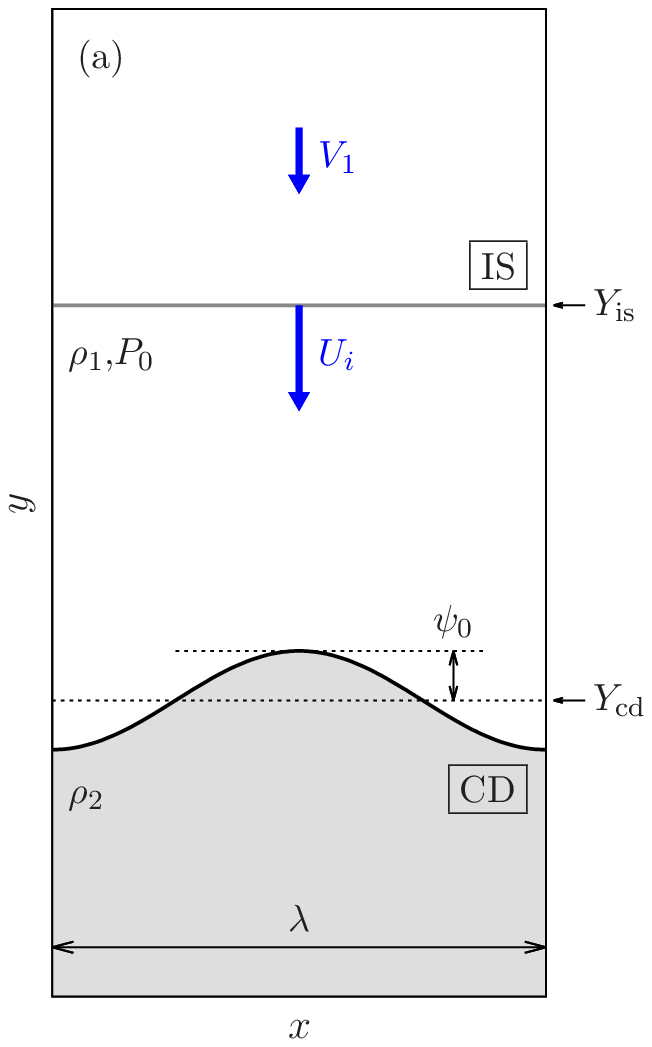}
\includegraphics[scale=0.95,clip]{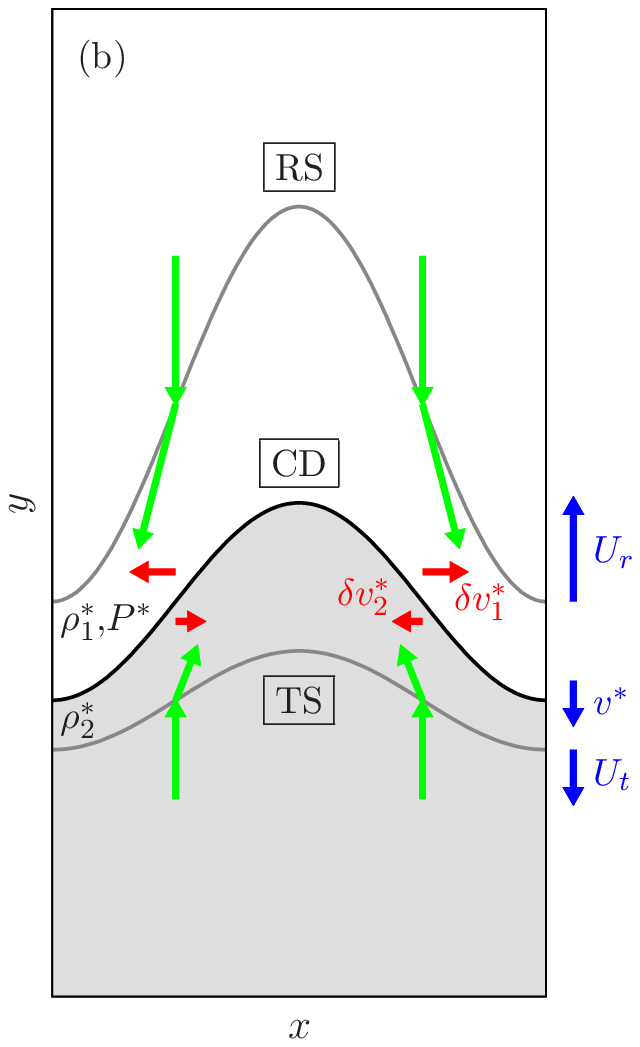}
\caption{
(a) 
Schematic picture of the initial configuration for single-mode
analysis of the RMI.   
Two fluids are divided by a contact discontinuity (CD). 
The densities of the light fluid ``1'' and heavy fluid ``2'' are
$\rho_1$ and $\rho_2$, and the uniform pressure for the both fluids is
$P_0$.
The interface is corrugated sinusoidally with the wavelength $\lambda$
and the amplitude $\psi_0$.  
An incident shock (IS) propagates in the light fluid ``1'' with the
shock velocity $U_i$.   
The shock strikes the corrugated interface at a time $t = 0$. 
Here $V_1$ is the flow velocity behind the incident shock.
(b) 
Sketch of the shock-front shapes after the incident shock hits the
corrugated interface. 
The transmitted shock (TS) and reflected shock (RS) travel from the
contact discontinuity in the opposite direction with the velocities
$U_t$ and $U_r$, respectively. 
The pressure and velocity at the contact discontinuity are $P^{\ast}$
and $v^{\ast}$, and the densities behind the transmitted and reflected  
shocks are $\rho^{\ast}_1$ and  $\rho^{\ast}_2$.
Because of the obliqueness of the shock surface, tangential flows,
$\delta v_1^{\ast}$ and $\delta v_2^{\ast}$, are generated at the both
side of the interface.
Refraction of the fluid motions at the transmitted and reflected
shocks are shown by the thick arrows.
\label{fig1}}
\end{center}
\end{figure}

The initial configuration for our single-mode analysis is illustrated
in Figure~\ref{fig1}$a$. 
Two fluids with different densities, $\rho_1$ and $\rho_2 (> \rho_1)$,
are separated by a contact discontinuity at $y = Y_{\rm cd}$. 
A shock propagating through the light fluid ``1'' ($y > Y_{\rm cd}$) 
hits the interface at a time $t = 0$. 
The incident shock velocity is $U_i (< 0)$ and the fluid velocity
behind the shock is $V_1 (< 0)$. 
Let $P_0$ be a uniform pressure of the both fluids before shock passage.
The Mach number of the incident shock is defined as $M = |U_i|/c_{s1}$
where $c_{s1} = (\gamma P_0/ \rho_1)^{1/2}$ is the sound speed of the
fluid ``1''. 
Here the $x$- and $y$-axis are set to be perpendicular and parallel to
the shock surface.  

The interface has an initial corrugation of a sinusoidal form, $y =
Y_{\rm cd} + \psi_0 \cos (k x)$, where $\psi_0$ is a corrugation
amplitude, $k = 2 \pi / \lambda$ is the perturbation wavenumber, and
$\lambda$ is the wavelength.   
This spatial corrugation of the interface is an essential ingredient
for the occurrence of the RMI.

The initial geometry of a magnetic field is assumed to be uniform with
the size of $|\mbox{\boldmath $B$}| = B_0$ in the preshocked region
($y < Y_{\rm is}$).  
As for the field direction, three cases are considered and those are
for the cases of perpendicular MHD shocks ($B_x = B_0$), parallel
shocks ($B_y = B_0$), and oblique shocks ($B_x = B_y$). 
The physical quantities in the postshocked region behind the incident
shock are calculated from the Rankine-Hugoniot conditions for MHD
shocks.  
%For example, when the Mach number is sufficiently large, the density
%of the fluid ``1'' is enhanced by a factor of $(\gamma + 1)/(\gamma -
%1) \approx 4$ due to shock compression.
%Then the tangential component of the field behind the incident shock
%is larger by the same factor than that in the preshocked region.  

The initial configuration depicted by Figure~\ref{fig1}$a$
can be characterized by only four non-dimensional parameters. 
The sonic Mach number $M$ parameterizes the incident shock velocity.
The contact discontinuity is expressed by the density jump
$\rho_2/\rho_1$ and the ratio of the corrugation amplitude to the
wavelength $\psi_0/\lambda$.  
The initial field strength is given by the plasma beta $\beta_0 = 8
\pi P_0/ B_0^2$ which is the ratio between the gas and magnetic pressures
defined at the preshocked region.
Thanks to the simplicity of the system, a variety of situations can be
examined only by choosing these 
four parameters; $M$, $\rho_2/\rho_1$, $\psi_0/\lambda$, and
$\beta_0$.  

\subsection{Characteristic Scales for the RMI}

Figure~\ref{fig1}$b$ is a schematic picture around the contact
discontinuity after the shock passage. 
Now, the contact discontinuity moves with the velocity $v^{\ast}$.
The pressure $P^{\ast}$ becomes higher compared to the initial $P_0$. 
The incident shock is transmitted into the heavy fluid ``2'' ($y <
Y_{\rm cd}$) which moves with the velocity $U_t (< 0)$, and the density
behind the transmitted shock is $\rho^{\ast}_2$. 
Similarly, a reflected shock forms with the velocity $U_r (> 0)$ and
compresses the fluid ``1'' to the density $\rho_1^{\ast}$. 
If the interface is corrugated, the shock fronts of these two waves
are also rippled.  
Then, refraction of the fluids across the shocks occurs as indicated
by the arrows in Figure~\ref{fig1}$b$.

Consider fluid motions near the shock surface in a frame moving with
the transmitted or reflected shock. 
For the both shocks, the upstream fluid moves along the $y$-axis.
However, the downstream motion should have the tangential component
as a consequence of the obliqueness of the shock front. 
When the corrugation amplitude is small, the velocities of these
tangential flows are given by  
\begin{equation}
\delta v_1^{\ast} = k \psi_r (v^{\ast} - V_1) \;,
%\delta v_1^{\ast} = k \psi_r (V_1 - v^{\ast}) \;,
\end{equation}
\begin{equation}
\delta v_2^{\ast} = k \psi_t v^{\ast} %\;,
%\delta v_2^{\ast} = - k \psi_t v^{\ast} %\;,
\end{equation}
\citep{wouchuk97}, where $\psi_r = \psi_0 (1 - U_r/U_i)$ and $\psi_t
= \psi_0 (1 - U_t/U_i)$ are the initial ripple amplitudes for the
reflected and transmitted shocks.  
The tangential shear motion across the contact discontinuity generates
the vorticity at the interface, and which is the driving source of
unstable growth for the RMI.  

In the RMI, the shock acceleration is impulsive and causes the
perturbation amplitude to grow linearly in time.
\citet{richtmyer60} has proposed a simple estimate of the growth
rate of the amplitude; $\partial \psi / \partial t = k \psi^{\ast}_0
A^{\ast} v^{\ast}$ 
where $\psi^{\ast}_0 = \psi_0 (1 - v^{\ast}/U_i)$ is the interface
amplitude just after shock passage and $A^{\ast} = (\rho_1^{\ast} -
\rho_2^{\ast})/(\rho_1^{\ast} + \rho_2^{\ast})$ is the
Atwood number of the postshocked interface.
This prescription is obtained by a generalization of the
Rayleigh-Taylor formula with an impulsive acceleration.

More complex analytic formulas for the asymptotic growth rate of RMI
are derived by \citet{wouchuk96,wouchuk97}.  
In the weak shock limit, the linear growth velocity $v_{\rm lin}$ can
be written using the tangential velocities $\delta v_1^{\ast}$ and
$\delta v_2^{\ast}$ as
\begin{equation}
v_{\rm lin} = 
\frac{\rho_1^{\ast} \delta v_1^{\ast} - \rho_2^{\ast} \delta v_2^{\ast}}
{\rho_1^{\ast} + \rho_2^{\ast}} \;,
\label{eq:vlin}
\end{equation}
and then the timescale of the RMI is characterized by
\begin{equation}
t_{\rm rm} = \frac1{k v_{\rm lin}} \;.
\label{eq:trm}
\end{equation}
Within the linear regime, the spike and bubble grow with the same
velocity $v_{\rm lin}$. 
For the weak shock cases, it is sufficient that the vorticity only at
the interface is considered for the derivation of $v_{\rm lin}$.
However, for the strong shock cases, the bulk vorticity left behind
the rippled transmitted shock cannot be ignored.
It is known that the bulk vorticity reduces the growth of the RMI. 
Therefore, in the strong shock limit, the growth velocity given by
Equation~(\ref{eq:vlin}) will be overestimate by a factor of a few
\citep{wouchuk97}.  
Furthermore the growth rate of the RMI could be reduced in the presence of
a magnetic field \citep{wheatley05,wheatley09}.
However, for sake of simplicity, we used Equations~(\ref{eq:vlin}) and
(\ref{eq:trm}) as the typical scales of the RMI throughout our
analysis in this paper. 

Once the initial conditions are set, the growth velocity 
$v_{\rm lin}$ can be derived by solving an appropriate MHD Riemann
problem. 
The analytic formula given by Equation~(\ref{eq:vlin}) intimates that
the faster growth can be realized by the larger amplitude $\psi_0$
and/or the larger Mach number $M$.  
When $M \gtrsim 2$, the growth velocity is roughly proportional to
$M$.
This means that the growth time of the RMI is quite different
depending on the incident shock strength.
In the limit of high density ratio, $\rho_2/\rho_1 \gg 1$, the growth
velocity is approximately estimated as $v_{\rm lin} \sim k \psi_0
|v^{\ast}| \sim k \psi_0 (\rho_2/\rho_1)^{-1/2} |V_1|$. 
However, within the range of $\rho_2 / \rho_1 \sim$ 1.5-100 considered
in our analysis, the growth time $t_{\rm rm}$ has little dependence on
the density jump.  

Just for reference, we estimate the characteristic quantities related
to the RMI occurring in the ISM.
The shock velocity of SNRs is 100-1000 km s$^{-1}$, so that the
typical Mach number is $M \sim$ 10-100 in the warm neutral medium. 
It is a common feature in the ISM that nonuniform structures consisted
of warm and cold neutral media.
The cold and dense filamentary clumps and the surrounding warm
diffuse gas are formed as a natural consequence of the thermal
instability \citep{field65,koyama00}.
The density ratio between the two phases is about 10-100.
The length of the filament ($\sim$ 1 pc) is roughly given by the most
unstable scale of the thermal instability, and the thickness of the
filament is about 0.1 pc \citep{inoue09}.
At the boundary between the two phases, there exists a transition layer,
the size of which is determined by the so-called Field length ($\sim$
0.01-0.1 pc).  
It is known that the transition layer is unstable under corrugational
deformations \citep{inoue06,stone09}.
The most unstable wavelength of this instability at the
evaporation/condensation front is typically 0.1 pc and the growth time
is approximately equal to the cooling time. 
Then we can assume the initial corrugation amplitude for the RMI is of
the order of the ratio of the transition thickness to the most
unstable wavelength, i.e., $\psi_0/\lambda \sim 0.1$. 

Using these typical quantities of the ISM, the growth velocity is
estimated as $v_{\rm lin} / c_{s1} \sim$ 1-10 and the timescale of
RMI is roughly given by $t_{\rm rm} / t_{s1} \sim$ 0.1-0.01, where
$c_{s1}$ is the sound speed in the diffuse gas and $t_{s1} =
\lambda/c_{s1}$ is the sound crossing time. 
Assuming $c_{s1} \sim 10$ km s$^{-1}$ and $\lambda \sim 0.1$ pc, these
values are corresponding to $v_{\rm lin} \sim$ 10-100 km s$^{-1}$ and
$t_{\rm rm} \sim$ 10$^2$-10$^3$ yr.
Then the growth time $t_{\rm rm}$ is much shorter than the cooling
time $t_{\rm cool}$ which is a few Myr in the typical ISM \citep{inoue09}. 
In the analysis below, we choose $M = 10$, $\rho_2/\rho_1 = 10$, and
$\psi_0/\lambda = 0.1$ as the parameters for a fiducial case. 

\subsection{Numerical Scheme and Grid}

We solve the ideal MHD equations by using a conservative Godunov-type
scheme \citep[e.g.,][]{sano98}.   
Operator splitting algorithm is adopted in our scheme.
The hydrodynamical part of the equations is solved by a second-order
Godunov method, using the exact solutions of a simplified MHD Riemann
problem.    
The one-dimensional Riemann solver is simplified by including only the
tangential component of a magnetic field.  
The characteristic velocity is then the magnetosonic wave alone, and
then the MHD Riemann problem can be solved in a way similar to the 
hydrodynamical one \citep{colella84}. 
The piecewise linear distributions of flow quantities are calculated
with a monotonicity constraint following van Leer's method
\citep{vanleer79}.   
The remaining terms, the induction equation and the magnetic tension
part of the equation of motion, are solved by the consistent MoC-CT
method \citep{stone92b,clarke96}, guaranteeing $\nabla \cdot
\mbox{\boldmath $B$} = 0$ within round-off error throughout the
calculation \citep{evans88}. 

The numerical scheme includes an additional numerical diffusion in the
direction tangential to the shock surface in order to care the
carbuncle instability \citep{hanawa08}.  
The numerical viscosity is added only in the regions where the
characteristics of either fast or slow wave converges, i.e., in the
regions potentially dangerous to the carbuncle instability. 
This treatment could be crucial in our analysis because this numerical
instability influences the size of baroclinic vorticity, which is
generated by the misalignment of the pressure gradient of the shock
and the local density gradient across the interface. 
%This treatment could be crucial in our analysis because this numerical
%instability influences the size of the vorticity generated from the
%baroclinic term.  

Calculations are carried out in a frame moving with the velocity
$v^{\ast}$ which is the interface velocity after the interaction with
the incident shock.
For convenience, we define the locus of $y = 0$ as where the incident
shock reaches the contact discontinuity at $t = 0$.
Then the postshocked interface stays at $y = 0$ if it is not
corrugated initially, or stable for the RMI.
The simulations start before the incident shock hits the corrugated
interface ($|Y_{\rm is} - Y_{\rm cd}| > \psi_0$), and thus the initial
time is negative $t_0 = - |Y_{\rm is} - Y_{\rm cd}| / |U_i|$.  

The system of equations is normalized by the initial density and sound
speed of the light fluid ``1'', $\rho_1 = 1$ and $c_{s1} = 1$, and the
wavelength of the density fluctuation $\lambda = 1$.
The sound crossing time is also unity in this unit, $t_{s1} = \lambda
/ c_{s1} = 1$.
Most of the calculations in this paper use a standard resolution of 
$\Delta_x$ = $\Delta_y$ = $\lambda/256$ unless otherwise stated.
A periodic boundary condition is used in the $x$-direction, and an
outflow boundary condition is adopted in the $y$-direction. 
The size of the computational box in the $x$-direction is always set
to be $L_x = \lambda$.
The $y$-length, on the other hand, is taken to be relatively larger,
$L_y \geq 10 \lambda$. 
The choice of $L_y$ is depending on the initial model parameters. 
All the calculations are stopped before the transmitted shock reaches
the edge of the computational domain. 

\section{Results}

\subsection{Magnetic Field Amplification}

\begin{figure}
\begin{center}
\includegraphics[scale=0.7,clip]{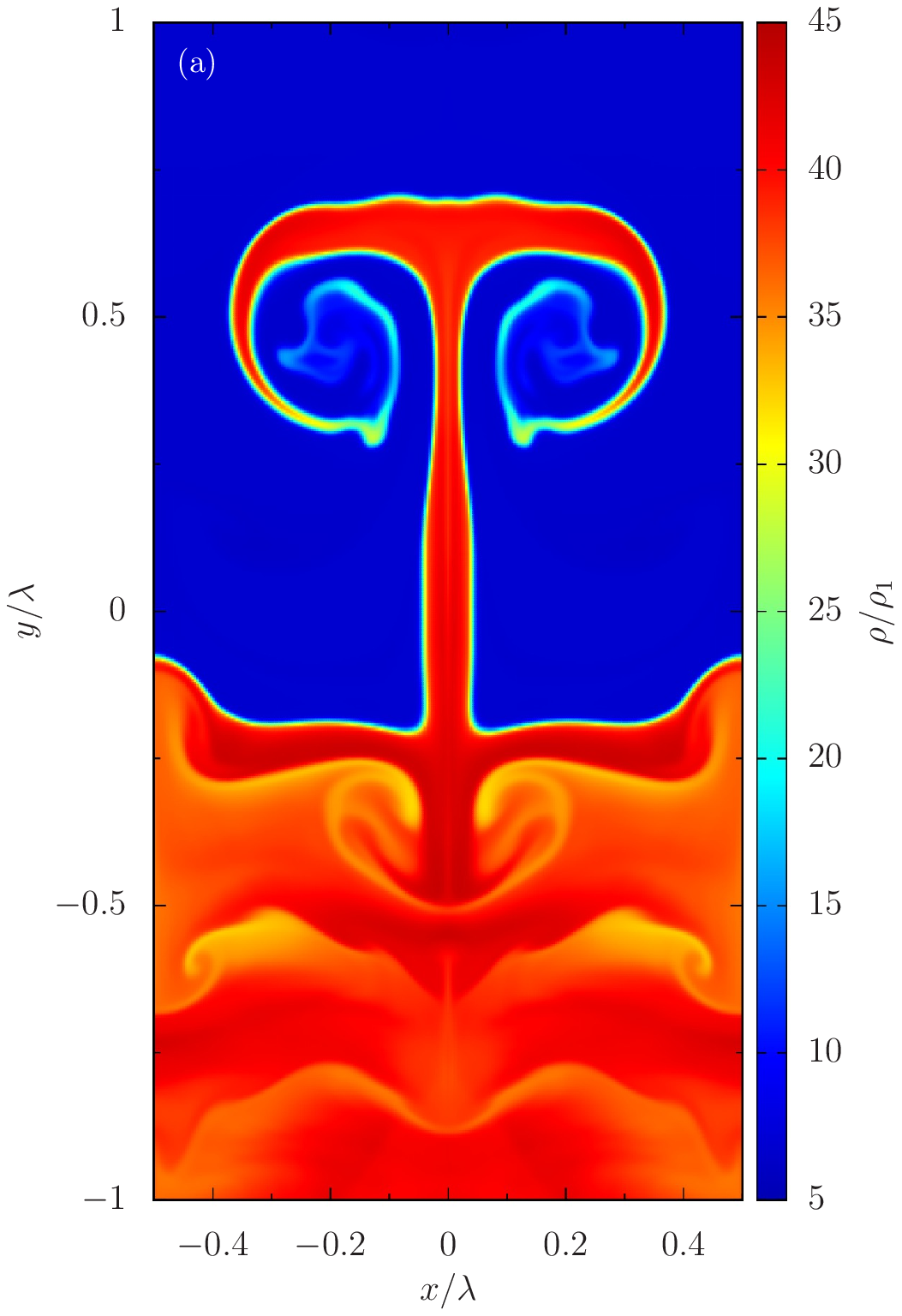}
\includegraphics[scale=0.7,clip]{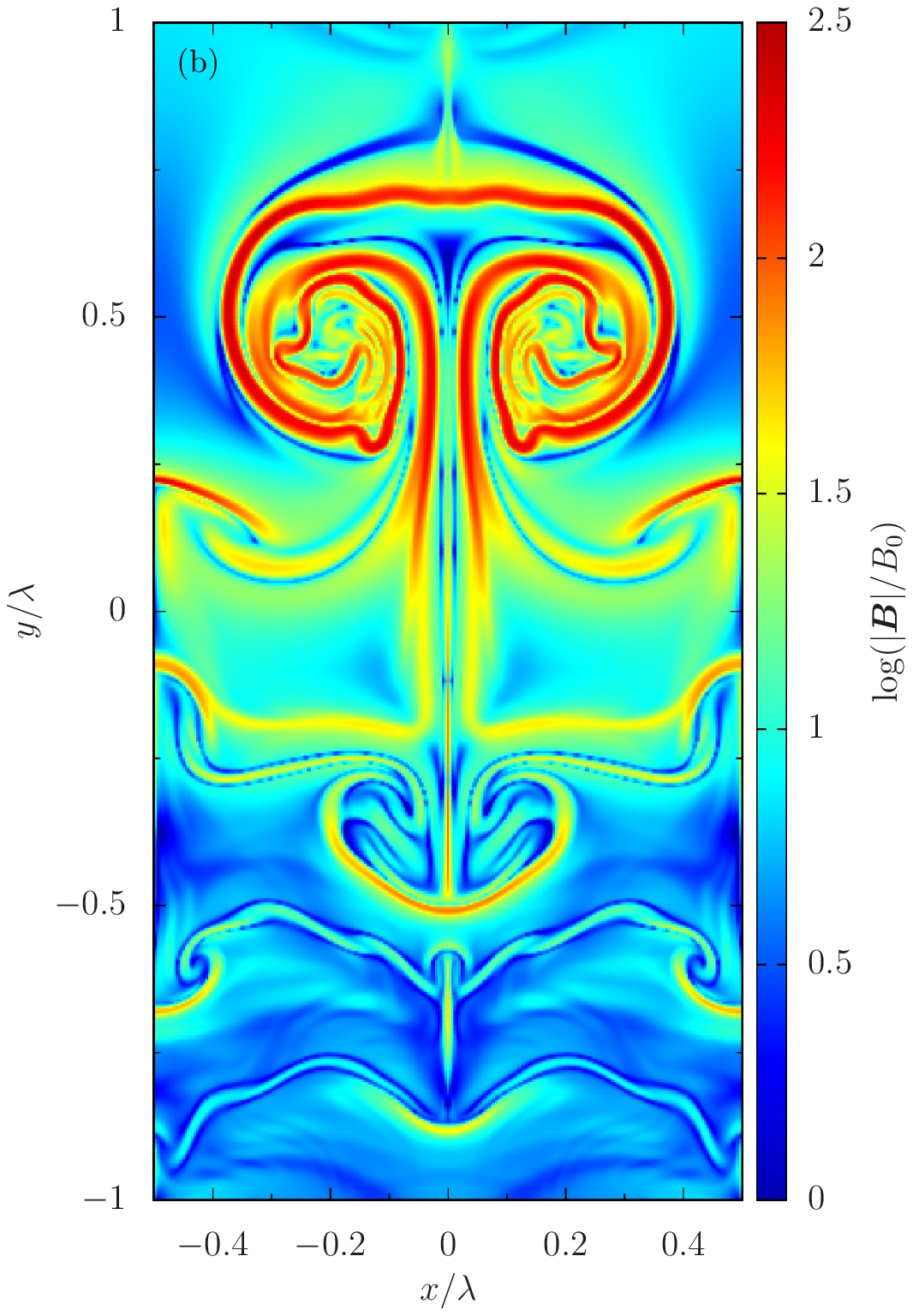}
\caption{
Spatial distributions of (a) the density and (b) the magnetic field
strength at a nonlinear stage of the RMI for the fiducial model.   
The model parameters are $M = 10$, $\rho_2/\rho_1 = 10$,
$\psi_0/\lambda = 0.1$, and $\beta_0 = 10^8$.  
The direction of the initial ambient field is in the $x$-direction, or 
perpendicular to the incident shock velocity (perpendicular MHD
shock). 
These snapshots are taken at the normalized time $k v_{\rm lin} t =
10$.
The maximum field strength at this time is
$|\mbox{\boldmath{$B$}}|_{\max} / B_0 = 272$. 
%A mushroom and spiral structures can be seen in the density profile as
%a result of the RMI. 
%Strong field regions are localized at the interface along the mushroom
%cap, where the ambient field is amplified largely by the fluid motions 
%associated with the RMI. 
\label{fig2}}
\end{center}
\end{figure}

Single-mode analysis of the RMI and the amplification of a magnetic field
are investigated by two-dimensional MHD simulations. 
Figure~\ref{fig2} shows the two-dimensional images of the density and
field strength at the nonlinear regime of RMI.
The initial parameters of this fiducial model are the Mach number $M =
10$, the density jump $\rho_2/\rho_1 = 10$, and the corrugation amplitude
$\psi_0/\lambda = 0.1$. 
A uniform magnetic field parallel to the shock surface, $(B_x, B_y) =
(B_0, 0)$, is assumed initially, and thus this case is for a
perpendicular MHD shock. 

In order to elucidate passive evolution of the field, the initial
strength is supposed to be very weak, $\beta_0 = 10^8$ at the upstream 
of the incident shock. 
%The initial field in the postshocked region ($y > Y_{\rm is}$) is
%amplified up to $B_x \approx 4 B_0$ by strong shock compression for
%this case.
Even though this weak field does not affect the dynamics of the RMI, the
spatial distribution of the field can be modified dramatically by the 
fluid motions driven by the RMI. 
Hereafter, we use the time normalized by the timescale of RMI,
$t_{\rm rm} = (k v_{\rm lin})^{-1}$. 
The ratio between $t_{\rm rm}$ and the sound crossing time $t_{s1}$
depends on the model parameters.
Most of the cases, $t_{\rm rm}$ is shorter than $t_{s1}$.
For example, $t_{\rm rm} = 0.116$ for the fiducial model, while
$t_{s1}$ is always unity in our simulations.
The snapshot data shown by Figure~\ref{fig2} are taken at the
normalized time $k v_{\rm lin} t = 10$. 

\begin{figure}
\begin{center}
\includegraphics[scale=0.95,clip]{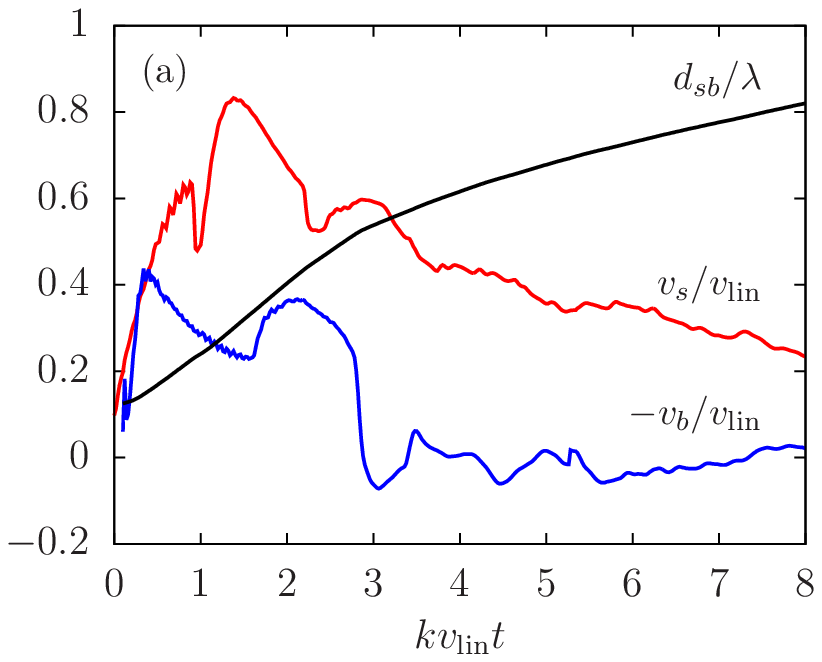}
\includegraphics[scale=0.95,clip]{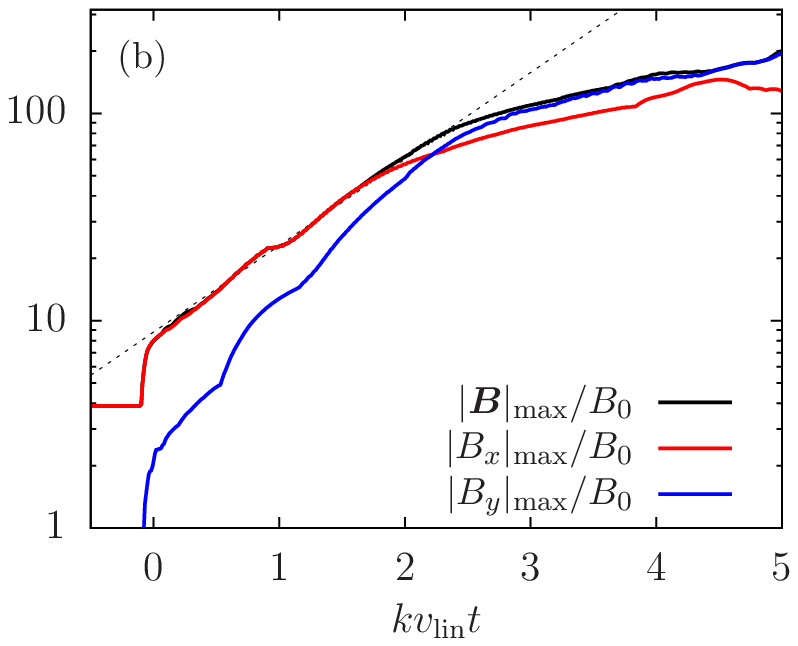}
\caption{
(a) 
Evolution of the growth velocities of the spike $v_s$ and the bubble
$v_b$, and the distance from the spike top to the bubble bottom
$d_{sb}$ in the fiducial model, which are normalized by the asymptotic
linear growth velocity $v_{\rm lin}$ and the wavelength of the
fluctuation $\lambda$. 
(b) 
Time profile of the maximum of the magnetic field strength
$|{\mbox{\boldmath$B$}}|_{\max}$ in the fiducial model.  
The maximum values of each component, $|B_x|_{\max}$ and
$|B_y|_{\max}$, are also shown.  
The exponential growth at the early stage until $k v_{\rm lin} t
= 2$ can be fitted by a function $\propto \exp (\sigma_B t / 
t_{\rm rm})$ with $\sigma_B = 1.0$ where $t_{\rm rm} = (k v_{\rm
  lin})^{-1}$ is the characteristic timescale of the RMI.   
\label{fig3}}
\end{center}
\end{figure}

The density profile shown in Figure~\ref{fig2}$a$ exhibits a
mushroom-shaped spike as a result of the growth of RMI.  
This figure is a close-up view focused around the contact
discontinuity at $y = 0$. 
The loci of the transmitted and reflected shocks at this time are
further out of this figure, that is, $y = -1.6 \lambda$ and
$5.6 \lambda$, respectively. 
The spike height reaches $y \sim 0.7 \lambda$, which is 7 times
larger than the initial amplitude $\psi_0$.  
In Figure~\ref{fig3}$a$, the time history of the distance from the
spike top to the bubble bottom $d_{sb} \equiv |y_s - y_b|$ is shown as
a function of the normalized time $kv_{\rm lin} t$, where $y_s$ and
$y_b$ are the $y$-coordinate of the spike and bubble.

Figure~\ref{fig3}$a$ also shows the growth velocities of the RMI,
which are evaluated by the advection velocities of the spike $v_s$ and
bubble $v_b$.
The growth velocity of the spike $v_s$ increases rapidly just after
the incident shock hits the interface, and takes the maximum when $k
v_{\rm lin} t \sim 2$. 
The maximum is comparable to the analytic prediction $v_{\rm lin}$
given by Equation~(\ref{eq:vlin}).
The growth velocity $v_{\rm lin}$ is subsonic compared to the sound
speed in the shocked regions, $v_{\rm lin} = 0.20 c_{s1}^{\ast} = 0.48
c_{s2}^{\ast}$ where $c_{s1}^{\ast} = (\gamma
P^{\ast}/\rho_1^{\ast})^{1/2}$ and $c_{s2}^{\ast} = (\gamma
P^{\ast}/\rho_2^{\ast})^{1/2}$, but it is supersonic relative to the
preshocked sound speed, $v_{\rm lin} = 1.37 c_{s1}$. 
The spike continues to grow with the velocity about $v_s \sim 0.2
v_{\rm lin}$ until at least $k v_{\rm lin} t = 30$ in this model.  

The bubble growth, on the other hand, is quenched at the earlier phase
of the evolution. 
This is a general feature of the RMI \citep{matsuoka03}, especially
for the strong shock cases as examined in this paper. 
The suppression is caused by the bulk vorticity comes from the
deformed transmitted shock.  
It is worth noticing that weak spike-like patterns can be seen in the
density distribution behind the transmitted shock in the heavy fluid
``2''.
This is also associated with the bulk vorticity generated by the
propagation of a rippled shock \citep[e.g.,][]{ishizaki97}.  
As the shock separates away from the interface, their corrugation
amplitude oscillates and decreases with time.
The amplitude of the bulk vorticity is larger for the cases with higher
shock strength, and it could affect the nonlinear behavior of the RMI
\citep{wouchuk97,nishihara10}. 
Although the effects of the bulk vorticity might be important for the 
interstellar strong shocks, its quantitative details is beyond the
scope of this paper. 

Figure~\ref{fig2}$b$ depicts the spatial distribution of the
field strength.  
The magnetic field is already amplified more than 200 times at $k
v_{\rm lin} t = 10$. 
The strong field regions are localized at the mushroom cap and form thin
filamentary structures along the interface. 
Figure~\ref{fig3}$b$ shows the time evolution of the maximum field
strength $|\mbox{\boldmath $B$}|_{\max}$ in the computational domain.
Before the incident shock hits the corrugated interface, the
$y$-component of the field is nothing in this model. 
The Mach number $M = 10$ is large enough that the density behind the
incident shock is enhanced by a factor of $(\gamma + 1)/(\gamma - 1)
\approx 4$ due to shock compression.  
Then the initial $|B_x|_{\max}$ is larger by the same factor than
$B_0$.  

During the interaction between the incident shock and interface, $B_y$
appears near the rippled interface due to the refraction of the flow. 
The maximum value $|B_y|_{\max}$ grows linearly in time at the very
beginning. 
Then each component of the field increases exponentially until $k
v_{\rm lin} t \sim 2$ during the growth velocity of the spike $v_s$ is
comparable to $v_{\rm lin}$ (see Figure~\ref{fig3}$a$).  
At this stage, $|B_x|_{\max}$ is still dominant over $|B_y|_{\max}$.
But the both components become comparable and evolve in a similar
manner at the later evolutionary stage $k v_{\rm lin} t \gtrsim 5$. 
The early growth rates $\sigma_B$ obtained by using a fitting formula,
\begin{equation}
|\mbox{\boldmath $B$}|_{\max} (t) \propto \exp 
\left( \frac{\sigma_B t}{t_{\rm rm}} \right) \;,
\end{equation}
are $\sigma_B = 1.0$, which is shown by the dotted line in the
figure. 
%and 1.4 for $|B_x|_{\max}$ and $|B_y|_{\max}$, respectively. 
The maximum strength increases with time even at $k v_{\rm lin} t = 30$
and exceeds $|\mbox{\boldmath $B$}|_{\max} \sim 10^3 B_0$ at the end.
These results are obviously suggesting that the RMI is a quite
efficient mechanism of magnetic field amplification. 

\subsection{Physical Mechanism of the Amplification}

In this subsection, we consider the physical mechanism of the field
amplification associated with the RMI.
The induction equation for the ideal MHD can be rewritten as 
\begin{equation}
\frac{\partial {\mbox{\boldmath{$B$}}}}{\partial t} = 
- ({\mbox{\boldmath$v$}} \cdot {\mbox{\boldmath$\nabla$}})
{\mbox{\boldmath$B$}}
+ ({\mbox{\boldmath$B$}} \cdot {\mbox{\boldmath$\nabla$}})
{\mbox{\boldmath$v$}}
- {\mbox{\boldmath$B$}} 
({\mbox{\boldmath$\nabla$}} \cdot {\mbox{\boldmath$v$}}) \; ,
\label{eqn:ind}
\end{equation}
where each term of the right-hand-side stands for advection,
stretching, and compression.
Among these terms, net increase of the field can be done only through
the stretching and compression.  

Figure~\ref{fig4}$a$ shows the magnetic field lines in the
neighborhood of the interface for the fiducial model at $k v_{\rm lin}
t = 2$. 
The gray color denotes the higher density part in the fluid ``2''
bounded between the contact discontinuity and transmitted shock front.  
As mentioned above, the magnetic field is amplified efficiently 
at the interface between two fluids.
The positions of the largest strength are shown by the crosses
in this figure. 
The fluid motion excited by the RMI stretches the surface area of the
contact discontinuity. 
It is found that the field amplification is predominantly originated
from this stretching effect.  

The relative importance of stretching and compression can be evaluated
from our numerical results. 
Multiplying the magnetic field vector $\mbox{\boldmath{$B$}}$ to
Equation (\ref{eqn:ind}),
the following scalar equation can be derived;
\begin{equation}
\frac12 \frac{\partial}{\partial t} | {\mbox{\boldmath$B$}} |^2 =
- {\mbox{\boldmath$B$}} \cdot ( {\mbox{\boldmath$v$}} \cdot
{\mbox{\boldmath$\nabla$}} ) {\mbox{\boldmath$B$}}
+ {\mbox{\boldmath$B$}} \cdot ( {\mbox{\boldmath$B$}} \cdot
{\mbox{\boldmath$\nabla$}} ) {\mbox{\boldmath$v$}}
- | {\mbox{\boldmath$B$}} |^2 {\mbox{\boldmath$\nabla$}} \cdot
{\mbox{\boldmath$v$}} \;,
\label{eq:sca}
\end{equation}
where the last two terms represent stretching and compression. 

Using the snapshot data same as in Figure~\ref{fig4}$a$, we calculate
the sizes of the stretching and compression terms in
Equation~(\ref{eq:sca}), which are shown in Figures~\ref{fig4}$b$ and
\ref{fig4}$c$, respectively.   
The same color contours are used in these two figures. 
Focusing on the mushroom-cap region, the stretching term takes the
maximum, while the compression term is almost nothing there. 
Because the fluid motions in the RMI is mostly
incompressible, so that the contribution of compression toward the
field amplification is negligibly small compared to that of stretching.   
The compression term is larger at the transmitted shock front, but
that is simply because of shock compression. 
The strong field regions along the interface perfectly match where
the stretching term is dominant over the other terms.
This is a clear evidence that the stretching effect is the major
source of the field amplification. 

To verify this interpretation, we compare quantitatively the
stretching rate of the interface and the growth rate of the field.
The stretching rate can be obtained numerically from a nonlinear
vortex sheet model \citep{matsuoka03}.
After the transmitted and reflected shocks have traveled a distance
larger than the fluctuation wavelength $\lambda$, the
system can be regarded as incompressible and irrotational except for
the interface on which nonuniform vorticity is induced by the shocks.
We treat the interface as a curve with the use of a Lagrangian
parameter in the $x$-$y$ plane.
The governing equations for the nonlinear analysis are the Bernoulli
equation and kinematic boundary conditions. 

The Bernoulli equation, i.e., the pressure continuous condition at the
interface, is given by 
\begin{eqnarray}
\rho_1^{\ast} \left[\frac{\partial \phi_1}{\partial t} +
  \frac{1}{2}(\nabla \phi_1)^2 \right] = 
\rho_2^{\ast} \left[\frac{\partial \phi_2}{\partial t} +
  \frac{1}{2}(\nabla \phi_2)^2 \right], \label{eqn:bernoulli}  
\end{eqnarray}
where $\phi_i$ $(i=1,2)$ is the velocity potential defined as $\nabla
\phi_i = \mbox{\boldmath $v$}_i$ and $\mbox{\boldmath $v$}_i$ is the
velocity of the fluid ``$i$''. 
In order to calculate the interfacial motion as a vortex sheet, we
need to rewrite Equation (\ref{eqn:bernoulli}) into an evolution
equation for the vortex sheet strength $\kappa$. Here,
$\kappa=|\mbox{\boldmath $\kappa$}|$ is defined by the circulation
$\Gamma \equiv \phi_1 - \phi_2$ as $\mbox{\boldmath $\kappa$} =
\nabla\Gamma$. The detailed derivation of the evolution equation
associated with the vortex-induced velocity is explained in the
appendix.  

The obtained shape of the interface at $k v_{\rm lin} t = 2$ is shown
by Figure~\ref{fig4}$d$.
The line color indicates the stretching rate $\sigma_{\rm int}$ at
each Lagrangian point on the interface, which is defined as 
\begin{equation}
\sigma_{\rm int} \equiv \frac{t_{\rm rm}}{l} \frac{dl}{dt} \;
\end{equation}
where $l$ is the line element along the interface. 
The vortex sheet result reproduces surprisingly well the spatial
distribution of the stretching size in our simulation shown by
Figure~\ref{fig4}$b$. 
The large stretching rate can be seen at the spike top, and
this feature coincides exactly with the strong field region as shown in
Figure~\ref{fig4}$a$.  
The vortex sheet model predicts that the largest stretching rate
appears at the side of the mushroom cap in the later phase.
This is also consistent with the MHD results at $k v_{\rm lin} t = 10$
depicted by Figure~\ref{fig2}$b$. 
Furthermore the maximum stretching rate $\sigma_{\rm int} \sim 0.8$ is
fairly close to the growth rate of the magnetic field $\sigma_B \sim
1.0$. 
This is another fact to convince us of the strong relation between
the field amplification and the stretching effect. 

\subsection{Dependence on the Initial Field Geometry}

The orientation of interstellar magnetic fields has no correlation
with the direction of supernova shocks.  
Therefore, it is interesting to examine the effects of the initial
field direction on the amplification process. 
A parallel shock case with the initial field $(B_x, B_y) = (0, B_0)$
and an oblique shock case with $(B_x, B_y) = (B_0/\sqrt{2},
B_0/\sqrt{2})$ are performed for this purpose.
The other parameters are identical to the fiducial model,
so that the magnetic field is again assumed to be very weak, $\beta_0
= 10^8$. 
The physical quantities in the downstream of the incident shock are
calculated from the jump conditions for MHD shocks, and the fast shock
condition is used for the oblique shock case. 

The field amplification due to the RMI is found to be independent of
the direction of the ambient field.   
Figure~\ref{fig5} shows the time evolutions of the maximum field
strength for the models with different initial field geometries. 
First, let us take a look at the early stage of the evolution until $k
v_{\rm lin} t \approx 10$.
For all the models, the magnetic field increases exponentially at
this stage and is amplified up to about 100 times as
large as the initial strength $B_0$.
The stretching term always contributes the field amplification
dominantly. 
The difference in the maximum strength at $t = 0$ is caused by the
compression of the tangential field $B_x$ behind the incident shock.  
For the strong shock limit, the ratio of the initial $|B_x|_{\max}$ in
our models should be (parallel) : (oblique) : (perpendicular) 
$= 1 : 2 \sqrt2 : 4$. 
The highest field at $k v_{\rm lin} t = 10$ is achieved in the
perpendicular shock case, but this is mainly because of the
difference in the initial $|\mbox{\boldmath{$B$}}|_{\max}$.
%Only the tangential component can be amplified by the shock
%compression before the RMI takes place.
The amplification factor from the initial maximum field is rather
similar for all these models. 
The reason is that the evolution of the RMI is unaffected by the weak 
field, and then the stretching motion is almost identical no matter
which direction the ambient field takes. 

Figures~\ref{fig5.5}$a$ and \ref{fig5.5}$b$ show the spatial profile
of the field strength and the field lines, respectively, for the
oblique shock case.   
These snapshots are taken at $k v_{\rm lin} t = 10$ the same as
Figure~\ref{fig2}. 
The density distribution is almost identical to that in the fiducial
model shown by Figure~\ref{fig2}$a$.  
Although asymmetric structures can be seen in the amplified
field because of the shock obliqueness, the strong-field regions are
concentrated along the interface at the mushroom-shaped spike.
The maximum field appears near the top of the spike and the strength
is $|\mbox{\boldmath{$B$}}|_{\max}/B_0 = 192$ at this time.
This result strongly supports that the field amplification in the
oblique shock case is also caused by the surface stretching of the
contact discontinuity as well as in the perpendicular shock case
(fiducial model). 

At the later evolutionary stage $10 \lesssim k v_{\rm lin} t \lesssim
30$, the maximum $|\mbox{\boldmath{$B$}}|$ keeps increasing with time
in the perpendicular and oblique shock cases (see Fig.~\ref{fig5}). 
When the fluid ``1'' has a tangential field component, $B_x$, the
total magnetic flux swept by the spike increases as long as the growth
of RMI continues. 
On the other hand, if there is only a normal component, $B_y$, the
magnetic flux involved by the RMI is fixed by the initial setting.
Then the size of the maximum field have to be saturated when all the
flux is confined in a sufficiently small region.
This is why the parallel shock case shows the saturation of the
growth, a flat plateau in the time-profile of 
$|\mbox{\boldmath $B$}|_{\max}$.
This interpretation implies the resolution effect on the amplification
factor. 
Actually the resolution dependence can be seen in some models with a
weak initial field, and this is discussed later in the next section.

The time evolutions of the average field strength
$|\mbox{\boldmath{$B$}}|_{\rm ave}$ are also shown in
Figure~\ref{fig5}. 
Here the average is taken over the regions that the pressure is larger
than $10 P_0$ in order to eliminate the preshocked regions. 
These time profiles of the maximum and average field are qualitatively
quite similar to those in the realistic interstellar simulations
[e.g., Fig. 5 in \citet{inoue12}]. 
This resemblance reminds that the RMI would be a fundamental process
of the field amplification in the ISM.
However, our results are still restricted within the two-dimensional
picture. 
Obviously three-dimensional consideration will be an important next
step for further understandings. 

We found some curious features in the field evolution for the parallel
shock case.   
Figures~\ref{fig6}$a$ and \ref{fig6}$b$ show snapshots of the field
lines and the size of the stretching term at $k v_{\rm lin} t = 6$. 
For this case, the position of the maximum field is quite different
from that in the other cases.
The field lines are concentrated at the stem of the mushroom rather
than the interface or the mushroom cap. 
The heavy fluid ``2'' near the interface moves horizontally toward the
$y$-axis just after the shock passage. 
The magnetic field is frozen into the fluid and thus it is gathered
near the $y$-axis by this motion.
The converging flow can escape toward the mushroom cap along the field
lines so that it can be realized as incompressible motion.
The largely amplified region coincides to where the stretching term,
$B_y (\partial v_y / \partial y) \approx - B_y (\partial v_x /
\partial x)$, is the highest.   
Interestingly, the magnetic field is amplified selectively in the
fluid ``2'' for the parallel shock cases. 
For the perpendicular shock cases, the velocity pattern is the same,
but the converging flow basically moves along the field lines and thus
no field amplification at the stem part.

\section{Discussions}

\subsection{Parameter Dependence}

In this section, we discuss the dependence of the field amplification
on the model parameters such as the Mach number $M$, and the density
jump $\rho_2/\rho_1$, and the initial field strength $\beta_0$.
It is found that the amplification of a magnetic field by the RMI
occurs in a variety of cases with a wide range of the initial
parameters. 
The amplification factor is more than 100 for most of the cases. 
Therefore, this process should be a quite common phenomena, especially
in astrophysical shock events with a high Mach number. 

Figure~\ref{fig7} shows the evolutions of the maximum field strength
for various models.
The dependence on the shock strength is shown in Figure~\ref{fig7}$a$.
The Mach number ranging from $M = 1.5$ to 100 are examined.
The model parameters except for $M$ are identical to the fiducial
model and the initial field orientation is in the $x$-direction.
When $M \gtrsim 3$, the amplification factor exceeds $10^3$ in $k
v_{\rm lin} t = 30$ as shown in the figure.
The time history of $|\mbox{\boldmath $B$}|_{\max}$ in terms of the
normalized time $k v_{\rm lin} t$ looks quite similar for all the
models.  
The amplification process seems to be independent of the Mach number
for this parameter range.
This is because the stretching rate of the interface in the normalized
unit has little dependence on the Mach number.
Thus, even for the weak shock case with $M=1.5$, the magnetic field
can be amplified by more than 100 times in $k v_{\rm lin} t \sim 10$.
These results are for the cases of the perpendicular shock, but the
same conclusion can be obtained for the parallel shock cases.
Notice that the growth velocity is nearly proportional to the Mach
number. 
The normalization of the time is $t_{\rm rm} = 1.17 t_{s1}$ for the
model with $M = 1.5$ and $t_{\rm rm} = 1.15 \times 10^{-2} t_{s1}$ for
$M = 100$.
The actual timescale of the RMI growth is
largely different depending on the shock strength.

Figure~\ref{fig7}$b$ demonstrates the dependence on the density jump
at the interface.
The models of the different density ratio, $\rho_2/\rho_1 = 100$,
10, 3, and 1.5 are shown in this figure.
The other parameters and initial field geometry are the same as in the
fiducial model. 
Again, the magnetic field is enhanced by more than two orders of
magnitude in $k v_{\rm lin} t \sim 10$ for all the models. 

The amplification factor has a positive correlation with
$\rho_2/\rho_1$.  
Since the growth velocity $v_{\rm lin}$ has little dependence on
$\rho_2/\rho_1$, the growth timescale is comparable for these cases;
$t_{\rm rm} = 0.40 t_{s1}$ for $\rho_2/\rho_1 = 1.5$ and $t_{\rm rm} =
0.19 t_{s1}$ for $\rho_2/\rho_1 = 100$. 
However, the nonlinear behavior of the spike is found to be quite
different at the later stage.
The spike height reaches $y_s \approx 2.6 \lambda$ at $k v_{\rm lin} t =
30$ for the model with $\rho_2/\rho_1 = 100$, while it is only $y_s
\approx 0.8 \lambda$ for $\rho_2/\rho_1 = 1.5$.
When the density ratio $\rho_2/\rho_1$ is larger, the total magnetic
flux swept by the spike is larger. 
This magnetic flux near the interface becomes a seed field to be
stretched by the RMI.
Then the maximum strength could be much higher 
as the ratio $\rho_2/\rho_1$ is larger.
%in the larger $\rho_2/\rho_1$ models.
For the parallel shock cases, on the other hand, this trend can not be
seen at all and the saturated amplitude is independent of
$\rho_2/\rho_1$. 

The time evolutions of the average field strength
$|\mbox{\boldmath{$B$}}|_{\rm ave}$ are also shown in Figure~\ref{fig7}. 
The average is taken over the postshocked regions swept by the incident
and transmitted shocks.
For all the models, the maximum strength is much larger than the
average.
The strong field regions are highly localized along the stretched
interface as a result of the RMI.

We performed some models that the ratio $\rho_2/\rho_1$ is less than
unity. 
Although the phase of the RMI becomes reversed in those models,
the magnetic field is amplified by many orders of magnitude in the
same manner. 
It should be noted that the growth velocity given by
Equation~(\ref{eq:vlin}) is applicable for the reflected rarefaction
wave. 
The phase inversion is expressed by the negative value of the
growth velocity.  
The spike can penetrate deeply into the light fluid even for the
rarefaction cases, and its structure does not change by much compared
with that in the reflection shock cases.  
The efficient field amplification occurs mostly near the interface of
the spike for the both cases.

\subsection{Saturation Level of Magnetic Field}

The dependence of the field amplification on the initial field
strength is shown by Figure~\ref{fig8}.
The maximum strength $|\mbox{\boldmath{$B$}}|_{\max}$ in this figure
is normalized by $(8 \pi P^{\ast})^{1/2}$ where $P^{\ast}$ is the
postshock pressure. 
The initial plasma beta of these models are $\beta_0 = 10^8$, $10^4$,
100, and 1.
The parameters except for $\beta_0$ are identical to the fiducial
model and the magnetic field is initially aligned along the
$x$-direction for all the models.
When $\beta_0$ is large enough, the initial field is amplified more
than two orders of magnitude.
However, if $\beta_0 \lesssim 100$, the amplification is limited; the
factor is $\sim 10$ for $\beta_0 = 100$ and only a few for $\beta_0 =
1$. 

When the initial strength of the ambient field becomes larger, the
saturation level at the nonlinear regime is almost converged
independent of $\beta_0$. 
The upper limit of the field strength is about $\beta^{\ast} = 8 \pi
P^{\ast} / |{\mbox{\boldmath{$B$}}}|_{\max}^2  \sim 10$, which is
roughly equal to the equipartition value to the thermal pressure after
the shock heating.  
More precisely, the saturated magnetic field strength is determined by
a balance with the kinetic energy of the growth velocity of the spike,
$|{\mbox{\boldmath{$B$}}}|_{\max}^2/8 \pi \approx \rho_2^{\ast} v_{\rm
  lin}^2 / 2$.
For the strong shock cases, both of them can be comparable to the
thermal energy. 
As seen from Figure~\ref{fig2}$b$, the amplified field distributes
along the interface at the mushroom cap with the thickness of $\sim 0.01
\lambda$-$0.02 \lambda$. 
If the magnetic pressure becomes comparable to the gas pressure, 
the interface stretching is reduced by the Lorentz force, and the
increase of the maximum field strength will be saturated.  
Then the potential limit of the field amplification is determined by a
condition that the RMI is suppressed by the amplified field itself.

In this figure, higher resolution results are also shown by the dotted
curves. 
The same colors denote the same initial conditions.
The resolution dependence can be seen for the models with a weaker
initial field, where the higher resolution gives the higher saturation
level.  
However, the upper limit of the saturated field has little dependence
on the grid size.
The parallel shock models shows qualitatively the same results for the
dependence on the initial $\beta_0$ and the grid resolution.
Therefore, we can conclude that the existence of the upper limit for the
saturated field is a robust nature. 

Even when the initial field strength is strong ($\beta_0 \sim 1$), the
magnetic field can be amplified by a huge factor if the Mach number is
much larger than about 10. 
For the strong shock cases, the postshock pressure $P^{\ast}$ is
larger than $P_0$ by many orders of magnitude.  
Then the magnetic pressure of the initial field, $B_0^2/8 \pi$, can be
regarded as rather weak in terms of $P^{\ast}$. 
Therefore, the vorticity is deposited very close to the interface,
and then the RMI can grow even in the presence of the strong field.
As a result of the nonlinear growth of the RMI, the magnetic field is
amplified up to the thermal value of the postshock pressure $P^{\ast}$.

The plasma beta in the ISM is usually close to unity
\citep{beck01,heiles05}.  
When $\beta_0 \sim 1$ initially, the upper limit of the
amplification factor is of the order of $|\mbox{\boldmath{$B$}}|_{\rm
  max} / B_0 \sim (\beta_0 P^{\ast}/P_0)^{1/2} \sim (P^{\ast}/P_0)^{1/2} $.
If $P^{\ast}/P_0 \gtrsim 10^4$, 100-fold enhancement of the magnetic
field can be expected. 
The postshock pressure $P^{\ast}$ is determined just by the Mach number
$M$ and density ratio $\rho_2/\rho_1$.
The pressure ratio $P^{\ast}/P_0$ is nearly proportional to the square
of the Mach number, but has little dependence on the density ratio.
When the Mach number is larger than $M \gtrsim M_{\rm crit} \approx 50$,
the shock heating enhances the pressure by more than four orders
of magnitude $P^{\ast}/P_0 \gtrsim 10^4$.
The Mach number of the shocks associated with SNRs is sufficiently
larger than this critical value. 
If the initial field strength is assumed to be about the average value
of the ISM, $\sim$ 5 $\mu$G, then the enhanced maximum
field reaches to the level of milligauss. 
Therefore, this mechanism could explain the origin of the strong field
spots observed in the downstream regions of the front shock of SNRs
\citep{uchiyama07,uchiyama08}. 

The critical Mach number $M_{\rm crit} \approx 50$ is independent of the
initial corrugation amplitude of the interface, but the size of
$\psi_0$ affects the growth velocity and timescale of the RMI.
Assuming $\psi_0/\lambda = 0.1$, the growth time $t_{\rm rm}$ is much
shorter than the sound crossing time, $t_{\rm rm} \lesssim 0.03
t_{s1}$, when $M \gtrsim 50$. 
Since the cooling time is comparable to $t_{s1}$, the magnetic field
can be amplified significantly before the decrease of the pressure by
the radiative cooling. 
The ionization fraction of the postshocked media is sufficiently large
enough that the non-ideal MHD effects, such as the ambipolar diffusion
and ohmic dissipation, are ignorable. 
Then the dissipation timescale of the amplified field is much longer
than the RMI timescale.

In our numerical analysis, a discontinuous density jump is assumed at
the interface. 
However, for the cases of the ISM, there must be a transition layer
with a finite thickness between two different media.
Then we also examined the effects of the transition layer on the
field amplification. 
The density profile of the layer is approximately given by a function
$\propto \tanh [(y-Y_{\rm cd})/L_{t}]$ where $L_{t}$ denotes the
thickness of transition layers.  
Although the growth rate of the magnetic field decreases with the
increase of the thickness $L_{t}$, the saturation level of the maximum
strength is found to be independent of $L_{t}$ if the transition layer
is thinner than the corrugation wavelength $L_{t} \lesssim \lambda$. 
Thus the field amplification by the RMI could be realized even in the
typical situations of the ISM.

\subsection{Observed Emissions and Particle Acceleration}

If the entire region of SNR RX J1713.7-3946 is magnetized up to $B
\sim 1$ mG, the origin of gamma-rays cannot be leptonic
\citep{uchiyama07}, which is inferred from the observed power ratio
of the synchrotron to inverse-Compton emissions \citep{aharonian07}.  
However, when the strongly magnetized regions is caused by the RMI
after the shock-cloud interaction, the averaged field strength of the
overall SNR would be much smaller than milligauss, depending on the 
volume filling factor of the cloud in the upstream ISM.   
Therefore, the origin of gamma-ray from RX J1713.7-3946 should be
discussed carefully taking account of other observational features,
such as the spectrum of the gamma-ray emission \citep{abdo11,inoue12}
and structure of surrounding ISM \citep{ellison10,fukui12}.

As we estimated in \S 2.2, the timescale of field amplification by the
RMI in young SNRs is about 100 yr, which indicates that the high-energy
electrons accelerated at the forward shock (or the transmitted shock)
by the diffusive shock acceleration are cooled down due to the
synchrotron loss before the field strength reaches milligauss-order.   
Thus, in order to explain observed synchrotron X-ray hot spots, an
additional accelerator other than the forward shock is necessary to
produce brightening of the synchrotron X-rays at the region of $B \sim
1$ mG.  
\citet{inoue12} discussed that the reflected shock waves induced by
the shock-cloud interactions can accelerate the electrons even in the
downstream region of the forward shock, and which could explain the
synchrotron emissions from strongly magnetized regions.

\section{Summary}

We have investigated the evolution of a magnetic field associated with
the RMI by using two-dimensional MHD simulations.
In terms of the field amplification, the importance of ``laminar
stretching'' driven by the RMI at the interface is successfully
demonstrated. 
In our single-mode analysis, an incident shock propagating through a
light fluid is considered to encounter a contact surface of a heavy fluid.
When the interface is spatially corrugated, the RMI takes place and a
mushroom-shaped structure develops in the density profile.
An ambient magnetic field is initially supposed to be uniform and
subthermal. 
Our numerical results for various situations suggest that the RMI is
an efficient mechanism of the amplification of the interstellar
magnetic fields.
The main conclusions are summarized below.

\begin{enumerate}
\item
The fluid motions associated with the RMI strengthen an ambient
magnetic field by many orders of magnitude.
This phenomenon can be seen in a wide range of the initial parameters.
The amplification factor is almost independent from the Mach number of
the incident shock
%, the density contrast at the interface, 
and the initial field direction, so that it could occur even for the
cases with a weak shock and/or a small density jump.
Therefore we can conclude that the RMI is a robust mechanism of the
ambient field amplification.  
\item
The physical mechanism of the field amplification is stretching 
associated with the nonlinear evolution of the RMI.
%by shear motions.
The magnetic field is amplified efficiently at where the stretching
term of the induction equation is predominant over the other terms.
In most cases, the strong field regions are localized along the
mushroom-shaped interface and form filamentary structures.
Curiously, only for the parallel shock cases, the maximum field
appears at the stem of the mushroom in the heavier fluid. 
\item
The amplified magnetic field is saturated when the magnetic pressure
becomes comparable to the thermal pressure after the shock heating.
This is because of the suppression of the RMI through the Lorentz force
of the amplified magnetic field. 
If the Mach number of the incident shock is larger than about 50, we
can expect at least more than 100-fold enhancement of the initial field.
Thus the RMI can be a promising origin of the interstellar strong
fields observed at the shock of SNRs \citep{uchiyama07}. 
\end{enumerate}

The MHD RMI would play an important role not only in other
astrophysical phenomena \citep[e.g.,][]{inoue11} but also in many
scientific fields such as interplanetary shocks \citep{wu03} and
inertial confinement fusion \citep{lindl92,holmes99}.
In this paper, the evolutions of an external magnetic field are
examined.
However, particularly for the case of laser plasmas, a self-generated
field often cannot be ignorable. 
The baroclinic term generates a magnetic field as well as the
vorticity.   
Then proper treatment of two-fluid effects should be included in
the analysis for that purpose, and which will be an interesting
subject for our future work. 

The RMI has been studied extensively by laboratory experiments
\citep[e.g.,][]{niederhaus03,chapman06}.
Laser plasmas can be a new platform to examine the RMI in
laboratories \citep{dimonte93} and the inclusion of the effects of a
magnetic field will be possible in such experiments
\citep{kuramitsu11}.   
Therefore the magnetic field amplification proposed in this paper
could be tested by laser experiments in the near future.  

\acknowledgments

We thank Shu-ichiro Inutsuka, Tomoyuki Hanawa, and Takahiro Kudoh for
useful discussions and comments. 
Computations were carried out on SX-8R at the Cybermedia Center and
SX-9/B at the Institute of Laser Engineering of Osaka University. 

\appendix
\section{Vortex Sheet Model}

Using the circulation $\Gamma=\phi_1 - \phi_2$ and average velocity
potential $\Phi$ defined by $\Phi = (\phi_1 + \phi_2)/2$, we rewrite
the Bernoulli equation (\ref{eqn:bernoulli}) as 
\begin{equation}
\frac{D\Gamma}{D t} = 2A^{\ast}\frac{D\Phi}{D t} - A^{\ast}\mbox{\boldmath
  $q$}\cdot \mbox{\boldmath $q$} + \frac{A^{\ast}+2{\tilde
    \alpha}}{4}\mbox{\boldmath $\kappa$}\cdot \mbox{\boldmath
  $\kappa$} - {\tilde \alpha} A^{\ast}\mbox{\boldmath
  $\kappa$}\cdot\mbox{\boldmath $q$} \;, 
\label{eqn:gamma} 
\end{equation}
in which the derivative $D/D t$ is given by
$$
\frac{D}{D t} = \frac{\partial}{\partial t} + {\mbox{\boldmath
    ${\bar v}$}}\cdot \nabla \;, \quad 
{\mbox{\boldmath ${\bar v}$}} =
\mbox{\boldmath $q$} + \frac{{\tilde \alpha}\mbox{\boldmath
    $\kappa$}}{2} \;, 
$$
where $\mbox{\boldmath $q$} = \nabla\Phi$, $\mbox{\boldmath $\kappa$}
= \nabla\Gamma$, $A^{\ast}$ is the Atwood number defined by $A^{\ast}
= (\rho_1^{\ast} - \rho_2^{\ast})/(\rho_1^{\ast} + \rho_2^{\ast})$,
and ${\tilde \alpha} = {\tilde \alpha}(A^{\ast})$ ($|{\tilde
  \alpha}|\leq 1$) is a weighting factor such that ${\tilde \alpha}
\ne 0$ for $A^{\ast} \ne 0$ \citep{matsuoka06}. 
Here, the spatial derivative is taken at the interface.  

We regard the interface in the RMI as a curve in the $x$-$y$ plane, and
parameterize it using a Lagrangian parameter $\theta$. The velocity of
the interface $(x,y)=\left(X(\theta, t),Y(\theta,t)\right)$ is derived as
\begin{equation}
X_t = U + \frac{{\tilde \alpha} X_\theta}{2 s_\theta}\kappa \;, \quad 
Y_t = V + \frac{{\tilde \alpha} Y_\theta}{2 s_\theta}\kappa %\;, 
\label{eqn:dXdY}
\end{equation}
\citep{baker82,matsuoka06}, where the vortex sheet strength $\kappa$ is
defined by $\kappa = 
{\partial \Gamma}/{\partial s} = {\Gamma_\theta}/{s_\theta}$, $s$ is arc
length of the sheet, and the subscript denotes the differentiation
with respect to the variable. The vortex-induced velocity $U = U (\theta,
t)$ and $V = V (\theta, t)$ are given by the Birkhoff-Rott equation:   
\begin{equation}
U(\theta, t) = -\frac{1}{4
  \pi}\int_{-\pi}^{\pi}\frac{\sinh\left[ Y(\theta, t) - Y(\theta',
  t)\right] \kappa(\theta', t)
  s_\theta(\theta')d\theta'}{\cosh\left[Y(\theta, t) - Y(\theta',
  t)\right]-\cos\left[X(\theta, t)-X(\theta', t)\right] + \delta^2} \;,
\end{equation}
\begin{equation}
V(\theta, t) = \frac{1}{4 \pi}\int_{-\pi}^{\pi}\frac{\sin\left[ X(\theta,
  t) - X(\theta', t)\right] \kappa(\theta', t)
  s_\theta(\theta')d\theta'}{\cosh\left[ Y(\theta, t) - Y(\theta',
  t)\right] - \cos \left[ X(\theta, t)-X(\theta', t)\right] +
  \delta^2} %\;, 
\label{eqn:UV} 
\end{equation}
\citep{birkhoff62,rott56}, where we regularize the Cauchy integral
using Krasny's $\delta$ \citep{krasny87}.  

Differentiating Equation (\ref{eqn:gamma}) with respect to $\theta$, we
obtain the following Fredholm integral equation of the second kind: 
\begin{equation}
\kappa_t = \frac{2A^{\ast}}{s_\theta}(X_\theta U_t + Y_\theta V_t) -
\frac{(1+{\tilde \alpha} A^{\ast})\kappa}{s_\theta^2}(x_\theta U_\theta +
y_\theta V_\theta) + \frac{A^{\ast} + {\tilde
    \alpha}}{4s_\theta}(\kappa^2)_\theta \;. 
\label{eqn:kappa} 
\end{equation}
Solving Equations (\ref{eqn:dXdY}) and (\ref{eqn:kappa})
simultaneously, we can determine the motion of a vortex sheet in the
RMI. 

\begin{figure}
\begin{center}
\includegraphics[scale=0.7,clip]{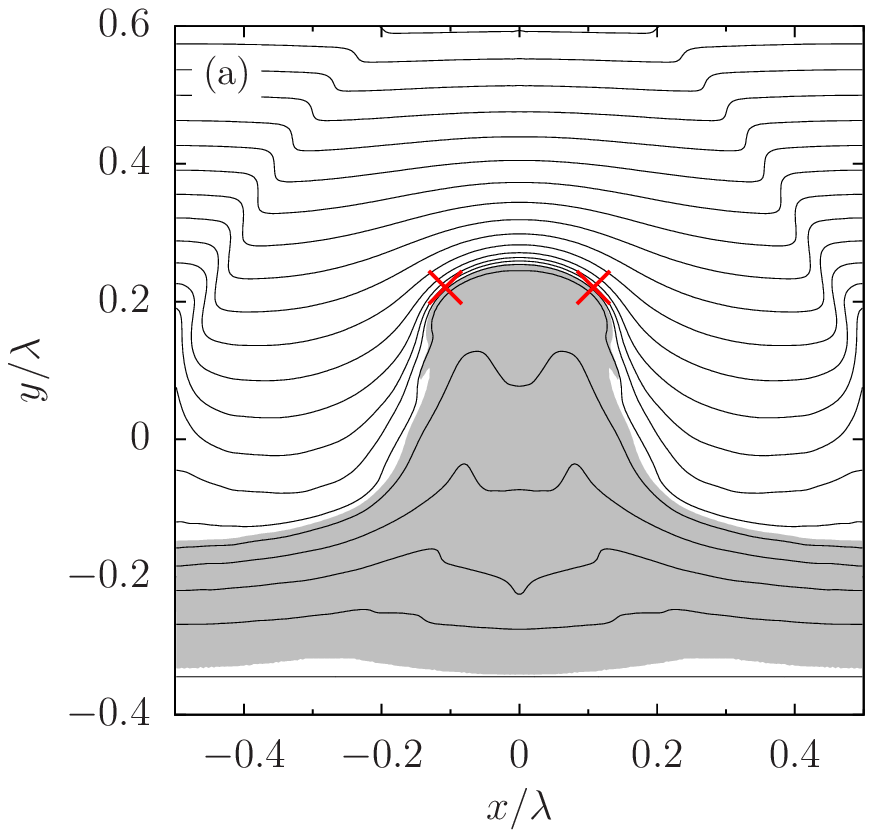}\hspace{1.55cm}
\raisebox{-0.1cm}{\includegraphics[scale=0.7,clip]{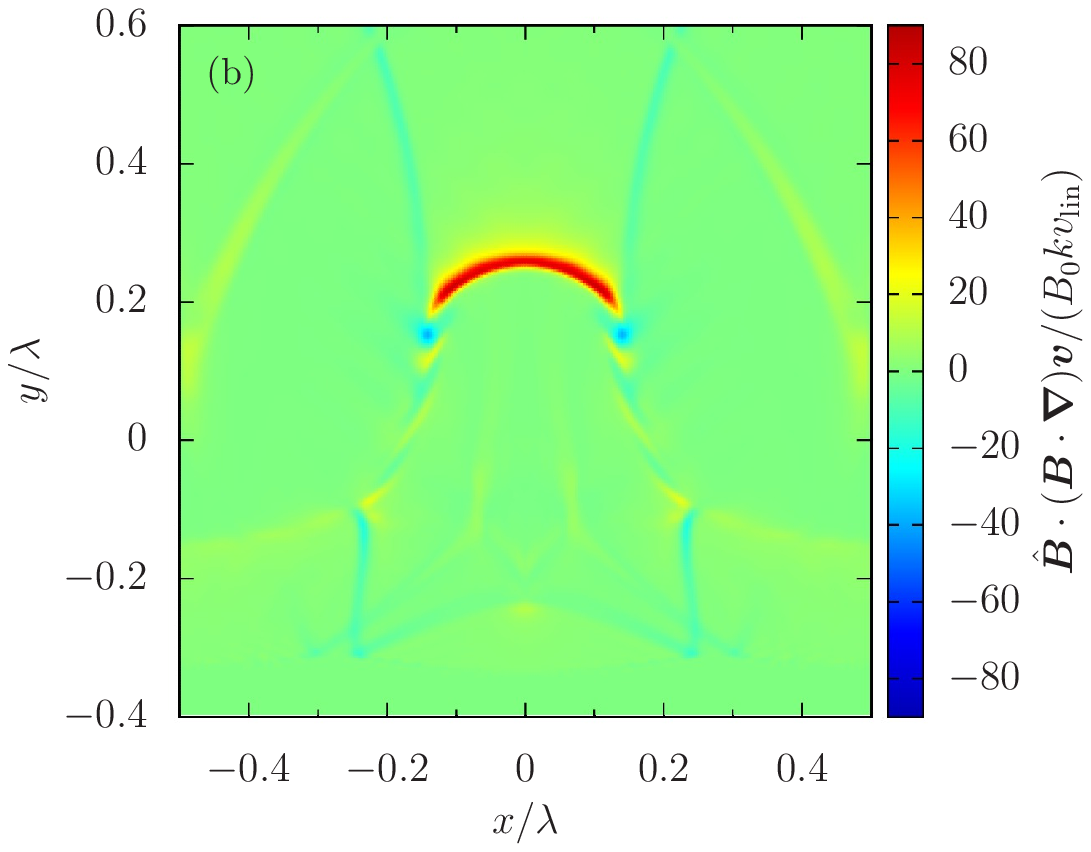}} \\
\includegraphics[scale=0.7,clip]{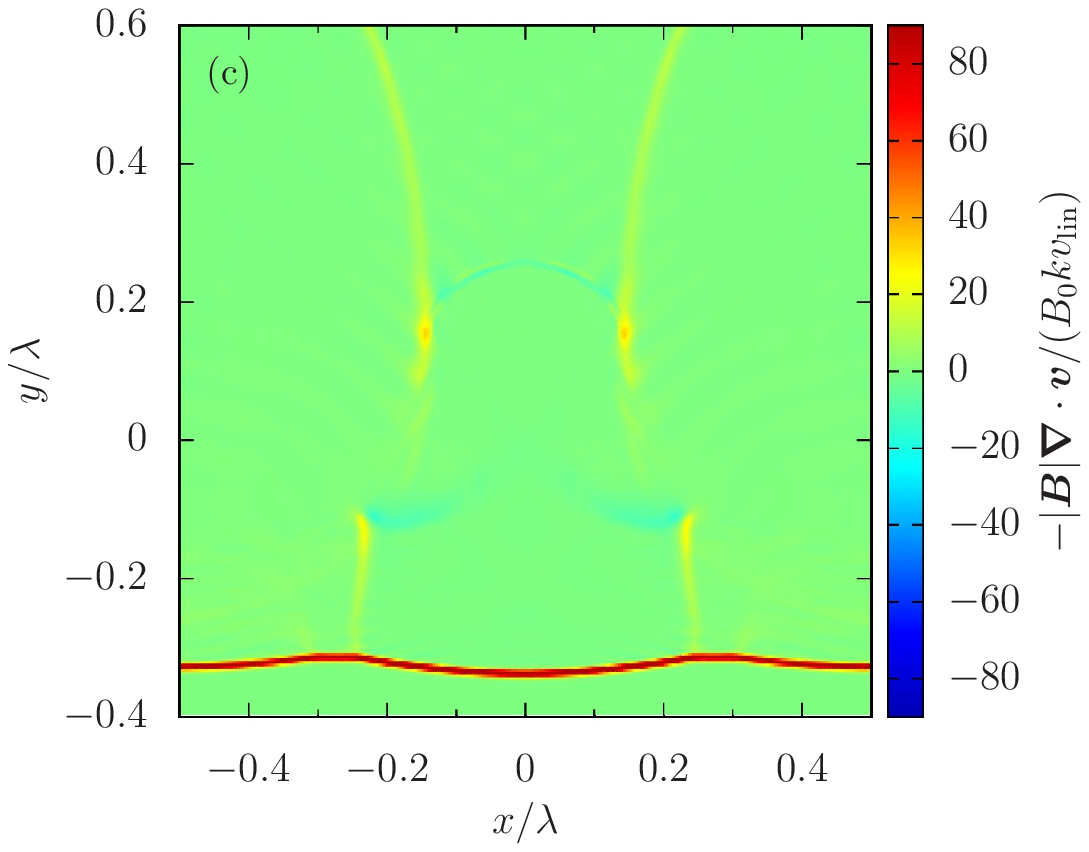}\hspace{0.1cm}
\raisebox{0.15cm}{\includegraphics[scale=0.7,clip]{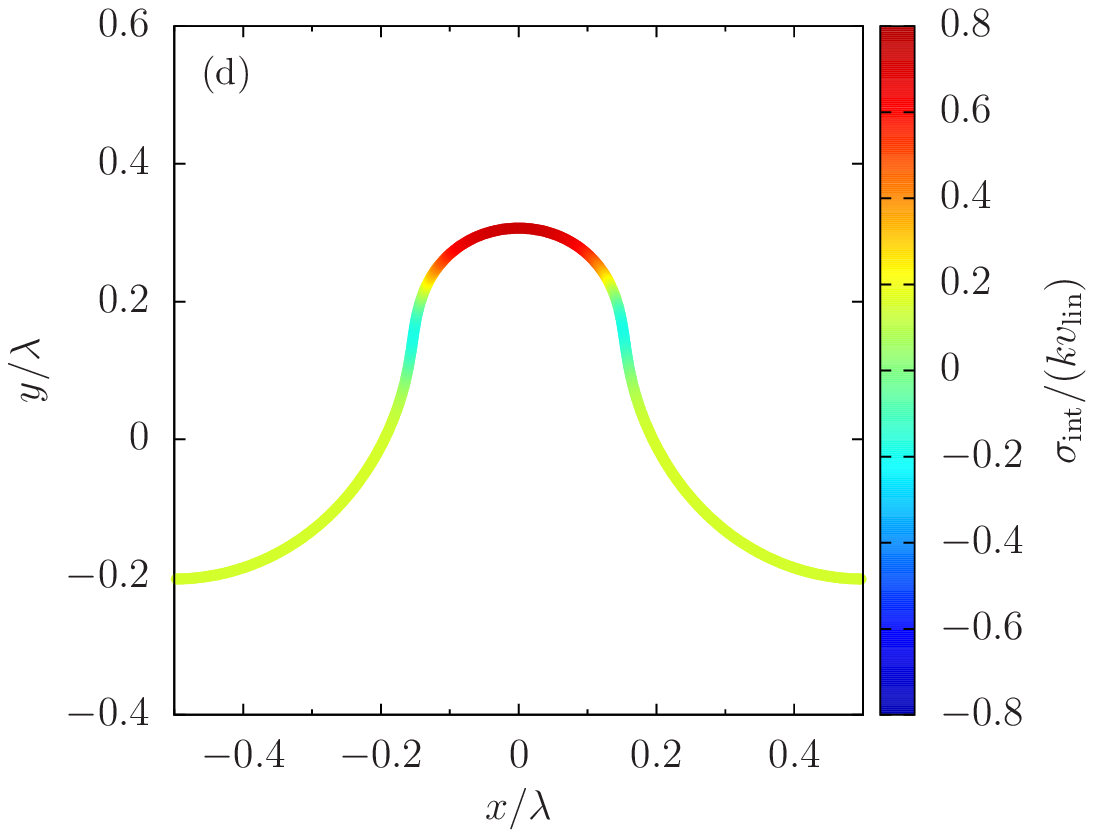}}
\caption{
(a) 
Magnetic field lines for the fiducial model at $k v_{\rm lin} t = 2$.
The gray color denotes the higher density regions in the fluid ``2''
compressed by the transmitted shock.
The positions of the maximum field are shown by the crosses. 
(b,c)
Relative importance of (b) ``stretching'' and (c) ``compression'' in the
induction equation (\ref{eq:sca}) calculated from snapshot data of the
fiducial model at $k v_{\rm lin} t = 2$.
Each term is normalized by a constant $B_0 k v_{\rm lin}$ and
$\hat{\mbox{\boldmath{$B$}}} \equiv \mbox{\boldmath{$B$}} /
|\mbox{\boldmath{$B$}}|$ is a unit vector. 
The same color bars are used for these figures.
(d)
Interface profile predicted by a vortex sheet model.
The model parameters corresponding to the fiducial model are used for
the numerical calculation.
The line color indicates the stretching rate of the interface
$\sigma_{\rm int}$, i.e., the rate of the temporal change of length at
each Lagrangian point.
\label{fig4}}
\end{center}
\end{figure}

\begin{figure}
\begin{center}
\includegraphics[scale=1.0,clip]{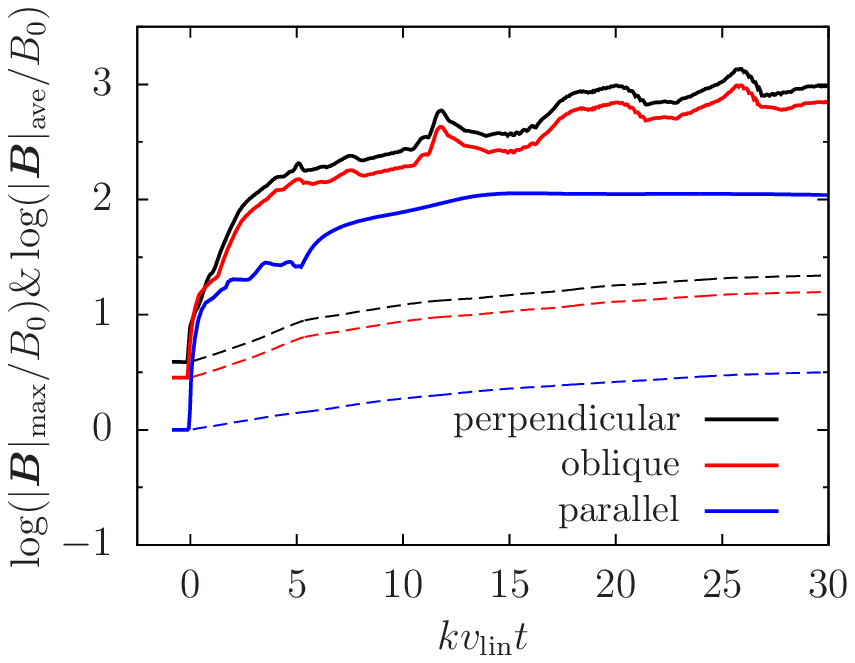}
\caption{
Time evolution of the maximum field strength
$|\mbox{\boldmath{$B$}}|_{\max}$ shown as a function of
the normalized time $k v_{\rm lin} t$ for the cases with different
orientations of the initial field.  
In the perpendicular shock case (fiducial model), the initial field is
$(B_x,B_y)=(B_0,0)$.
For the parallel shock case and the oblique shock case,
the uniform field direction is assumed to be
$(B_x,B_y)=(0,B_0)$ and $(B_x,B_y)=(B_0/\sqrt2,B_0/\sqrt2)$,
respectively. 
The other parameters are the same as in the fiducial model.
The thin dashed curves are time evolution of the average field strength
$|\mbox{\boldmath{$B$}}|_{\rm ave}$ for each case. 
The average is taken over only the postshocked regions.
\label{fig5}}
\end{center}
\end{figure}

\begin{figure}
\begin{center}
\includegraphics[scale=0.7,clip]{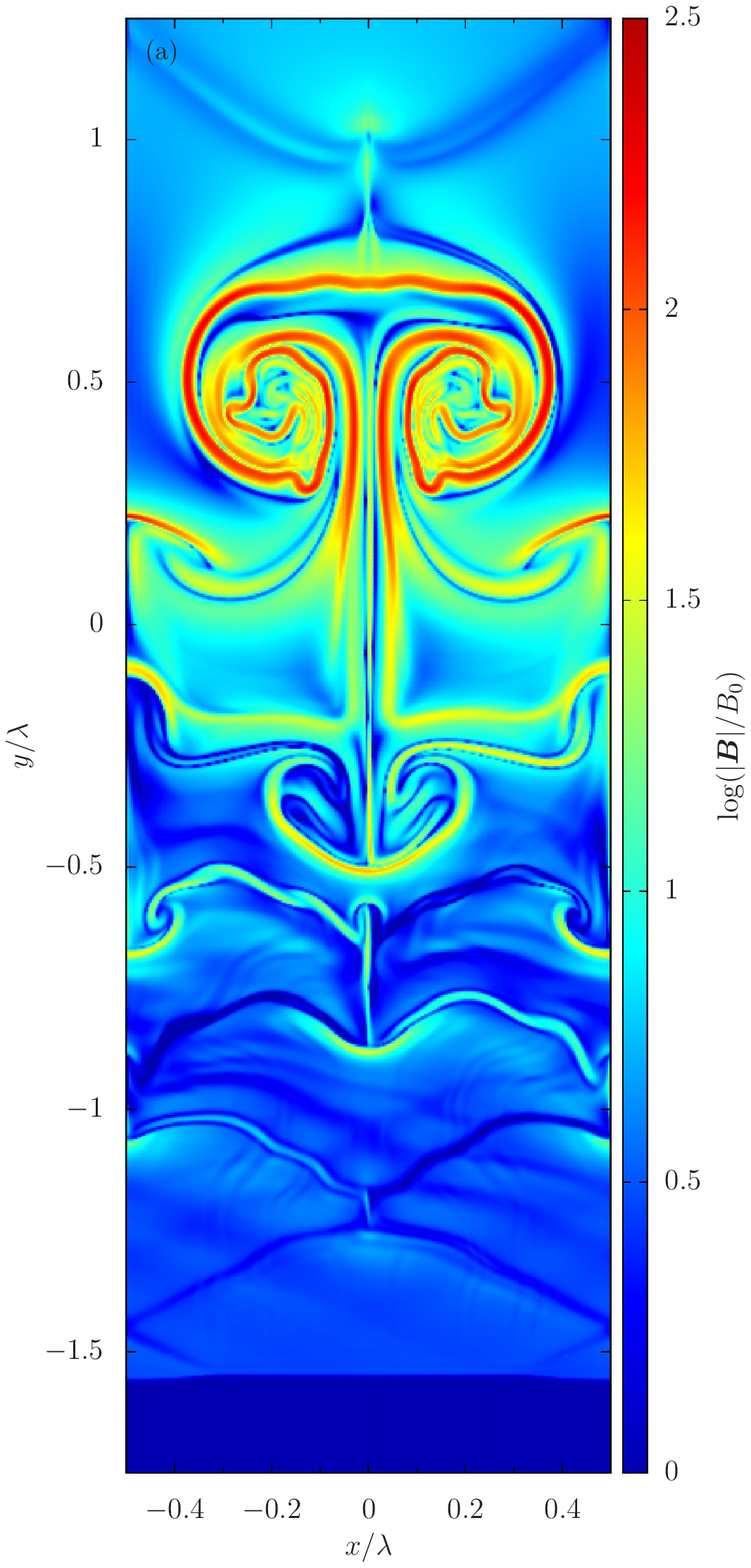}\hspace{0.1cm}
\raisebox{0.05cm}{\includegraphics[scale=0.7,clip]{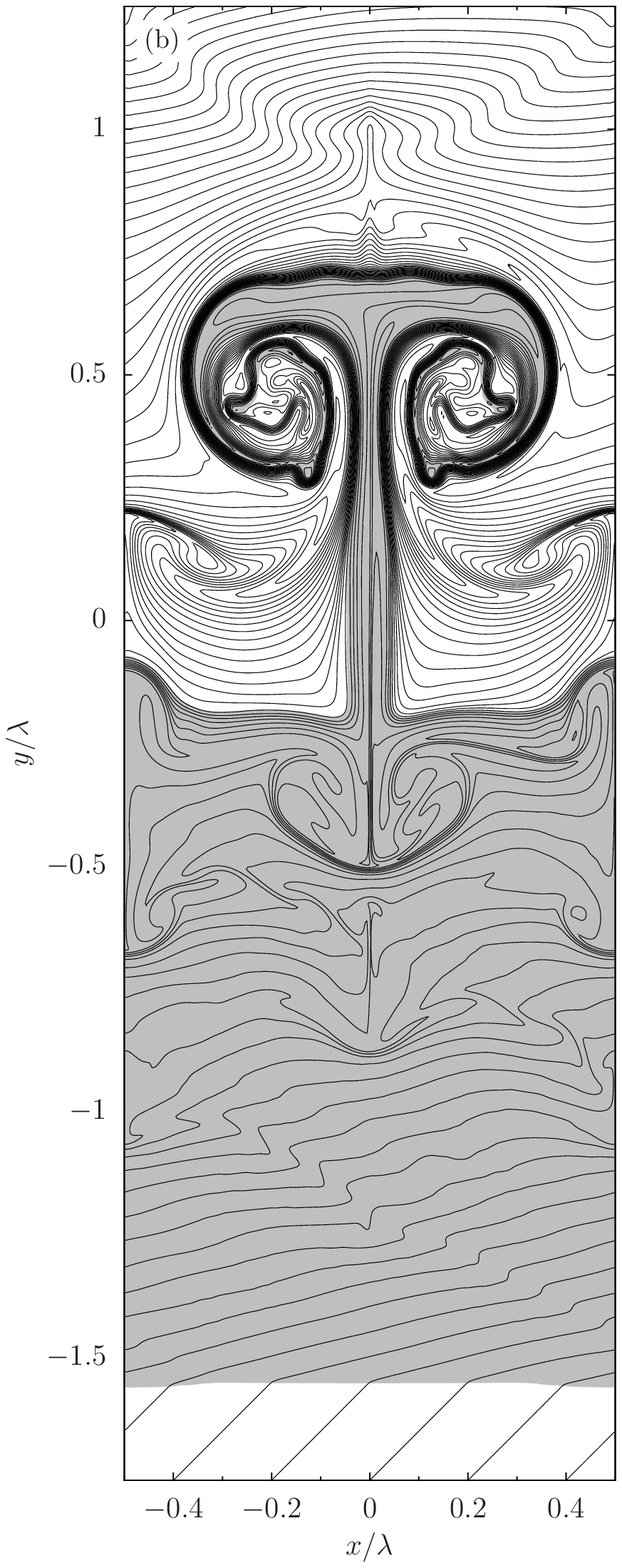}}
\caption{
Spatial distributions of (a) the magnetic field strength and (b) the
field lines for the oblique shock case.
The direction of the initial ambient field is 45-degree to the
$x$-axis.
The other parameters are identical to those in the fiducial model.  
These snapshots are taken at the normalized time $k v_{\rm lin} t =
10$. 
The higher density regions bounded between the contact discontinuity
and the transmitted shock front are depicted by the gray color in (b).
%The maximum field strength at this time is
%$|\mbox{\boldmath{$B$}}|_{\max} / B_0 = 192$.
\label{fig5.5}}
\end{center}
\end{figure}

\begin{figure}
\begin{center}
\includegraphics[scale=0.7,clip]{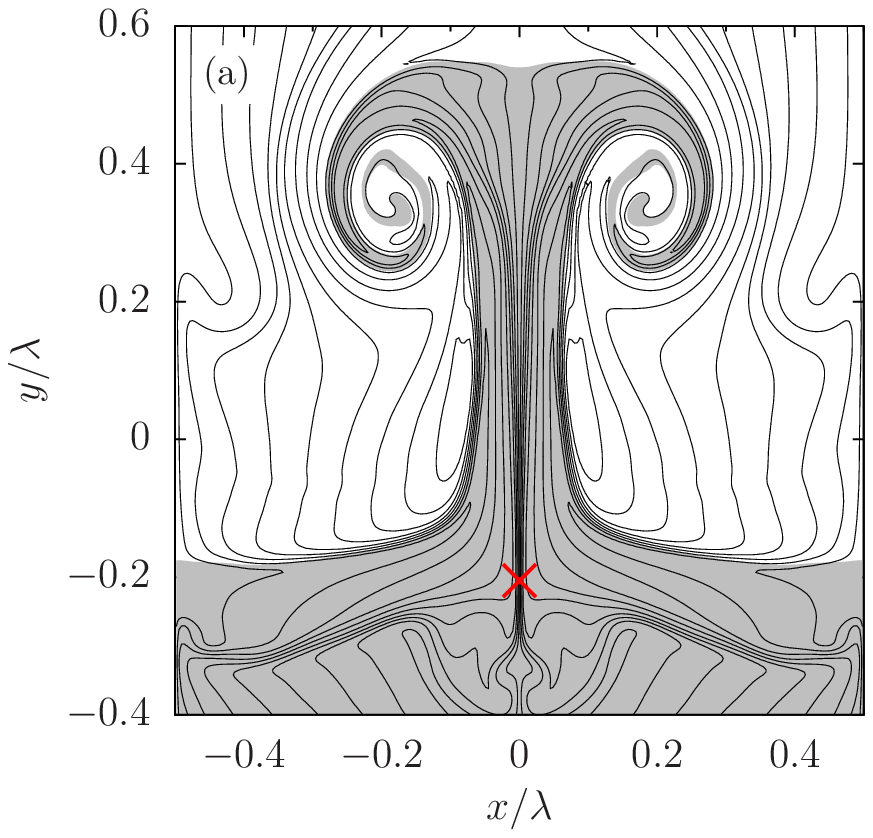}\hspace{0.1cm}
\raisebox{-0.1cm}{\includegraphics[scale=0.7,clip]{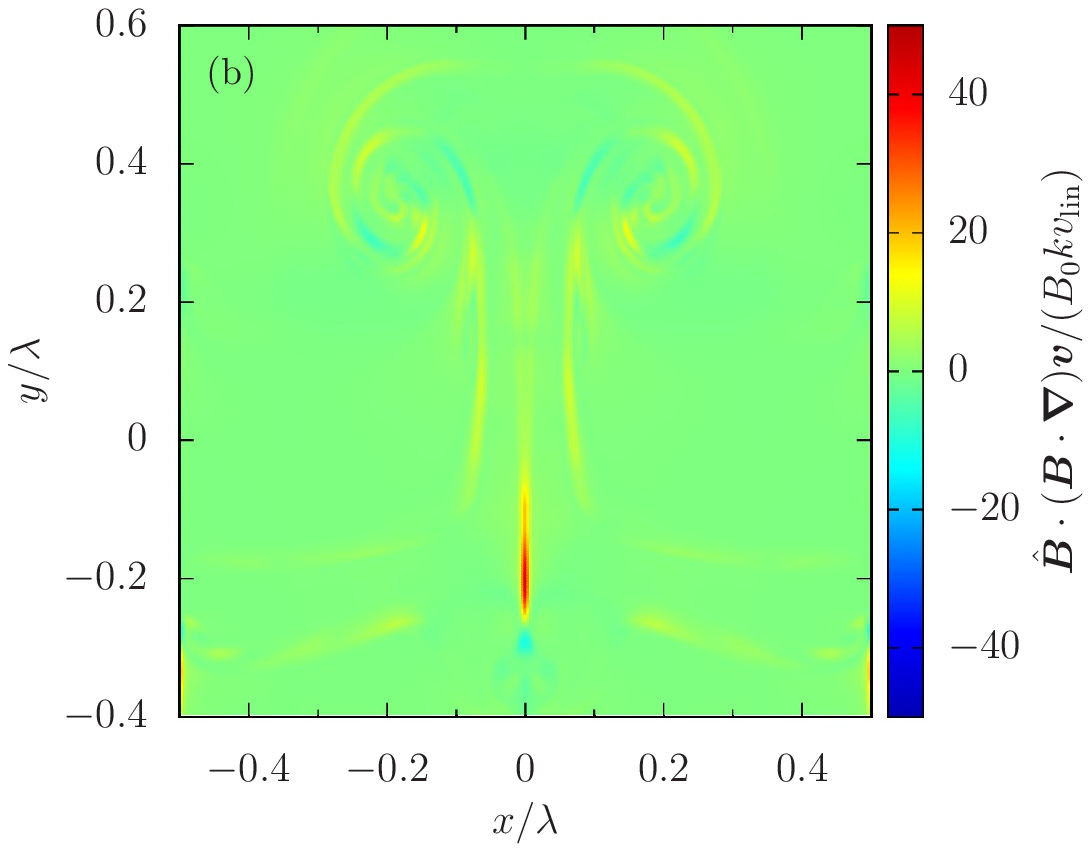}}
\caption{
(a)
Magnetic field lines of the parallel MHD shock case at $k v_{\rm lin} t =
6$.
The model parameters are identical to the fiducial model, but the
initial field direction is parallel to the incident shock
velocity for this case. 
The gray color denotes the higher density regions in the fluid ``2''
compressed by the transmitted shock.
The position of the maximum field is shown by the cross mark. 
(b)
Distribution of the size of ``stretching'' term in the induction
equation (\ref{eq:sca}) calculated from the snapshot data same as in
Figure~\ref{fig6}$a$.   
The stretching term is normalized by the same constant $B_0 k 
v_{\rm lin}$ as in Figure~\ref{fig4}$b$.
\label{fig6}}
\end{center}
\end{figure}

\begin{figure}
\begin{center}
\includegraphics[scale=0.95,clip]{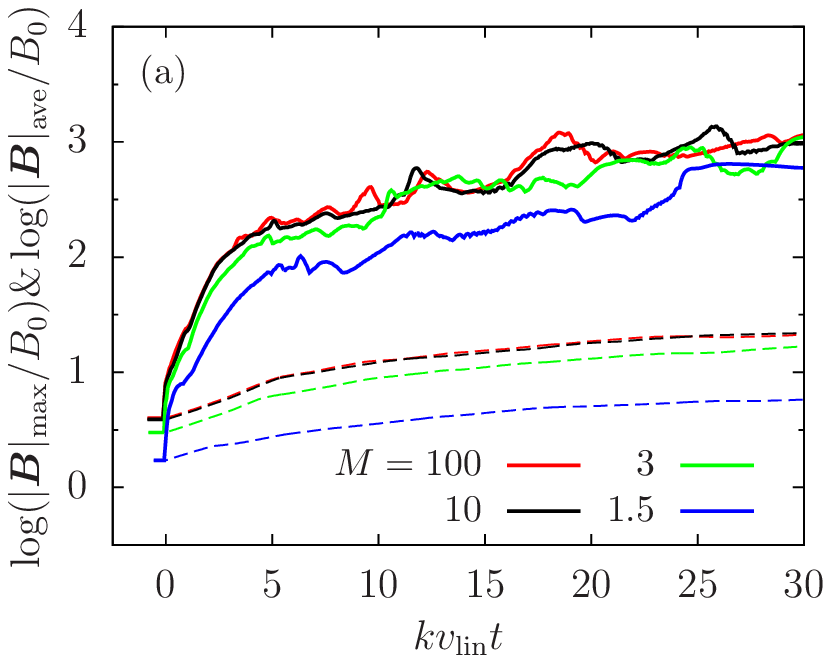}
\includegraphics[scale=0.95,clip]{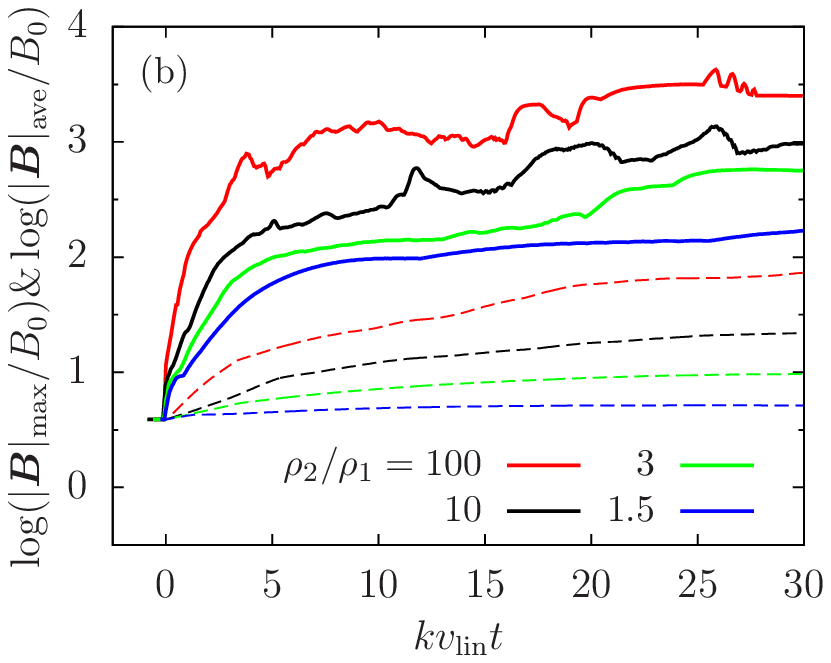}
\caption{
(a)
Dependence of the maximum field strength on the Mach number $M$.
The time profiles of $|\mbox{\boldmath{$B$}}|_{\max}$ are shown as a
function of the normalized time $k v_{\rm lin} t$ for the models with
different shock strength; $M=100$, $M=10$ (fiducial model), $M=3$, and
$M=1.5$.
The model parameters except for $M$ are identical to those of the
fiducial model. 
The thin curves are time evolution of the average field strength
$|\mbox{\boldmath{$B$}}|_{\rm ave}$ for each case.  
The average is taken over only the postshocked regions.
(b)
Dependence of the maximum field strength on the density jump
$\rho_2/\rho_1$. 
The time evolution of $|\mbox{\boldmath{$B$}}|_{\max}$ are drawn for the
models with different density ratio; $\rho_2/\rho_1=100$,
$\rho_2/\rho_1=10$ (fiducial model), $\rho_2/\rho_1=3$, and
$\rho_2/\rho_1=1.5$.
Other than $\rho_2/\rho_1$, the same parameters as in the fiducial
model are used.
The thin curves are time evolution of the average field strength
$|\mbox{\boldmath{$B$}}|_{\rm ave}$ for each case.  
\label{fig7}}
\end{center}
\end{figure}

\begin{figure}
\begin{center}
\includegraphics[scale=1.0,clip]{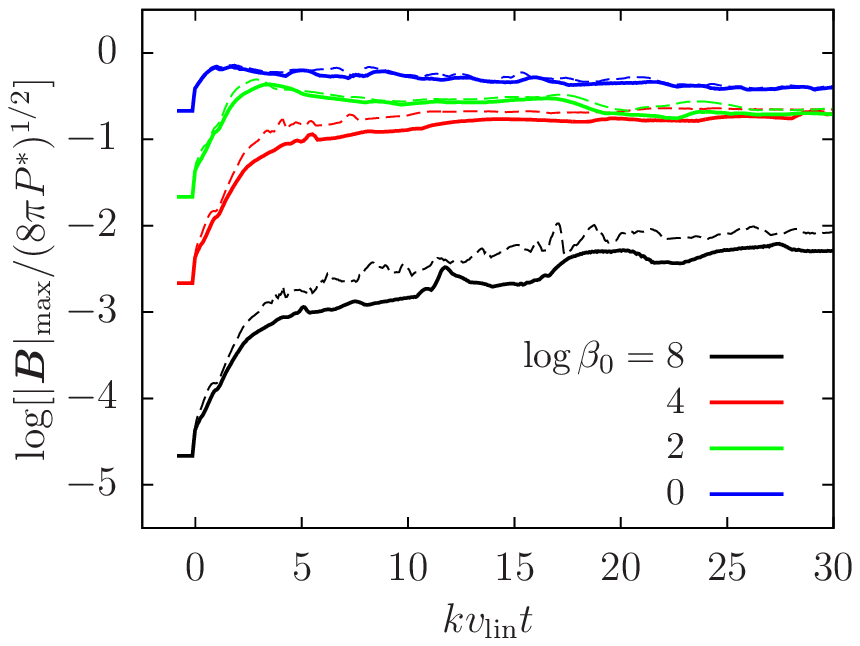}
\caption{
Dependence of the saturation level of the amplified magnetic field on
the initial plasma beta $\beta_0 = 8 \pi P_0 / B_0^2$.
The maximum field strength $|\mbox{\boldmath{$B$}}|_{\max}$ is
normalized by $(8 \pi P^{\ast})^{1/2}$ using the postshock pressure
$P^{\ast}$.  
The model parameters are the same as in the fiducial model except for
the initial field strength $B_0$, which is indicated by the line
colors; $\beta_0 = 10^8$ (fiducial model), $\beta_0 = 10^4$, $\beta_0
= 10^2$, and $\beta_0 = 1$. 
The initial field direction is assumed to be in the $x$-direction for
all the models. 
For the purpose of comparison, higher resolution results are also
shown in this figure by the thin dashed curves.
The grid size of the higher resolution runs is $\Delta_x = \Delta_y =
\lambda/512$ which is twice as good as the standard resolution.
\label{fig8}}
\end{center}
\end{figure}

For the fiducial model, the fluctuation amplitude after shock passed
is approximately calculated as $\psi_0^{\ast}/\lambda = 0.06188$. 
The density for the both layers are $\rho_1^{\ast}/\rho_1 = 6.716$ and
$\rho_2^{\ast}/\rho_1 = 39.549$, so that the Atwood number is
$A^{\ast} = - 0.7097$ for this case.   
The initial condition in Figure~\ref{fig4}$d$ is then given by
$$
X(\theta,0) = \theta \;, \quad 
Y(\theta,0) = a_0\cos \theta \;, \quad 
\kappa(\theta,0) = -\frac{2\sin\theta}{s_\theta(0)} \;, 
$$
where the initial amplitude $a_0$ is set to $a_0 = 2\pi \psi_0^{\ast}
/ \lambda = 0.3888$. 

In the numerical calculation, the stretching rate, i.e., the rate of
the temporal change of length $s(\theta)$ at the Lagrangian point
$\theta$ is defined by 
\begin{equation}
\frac{1}{{\bar s}(\theta)}\frac{ds(\theta)}{dt}
= \frac{1}{{\bar s}(\theta)} \frac{s(\theta,t+\Delta t) -
  s(\theta,t)}{\Delta t} \;,
\end{equation}
where ${\bar s} = \left[s(t+\Delta t) + s(t)\right]/2$ and the length
$s(\theta)$ is calculated by $\int s_\theta d\theta$, in which the
integral is performed as the spectral integration with respect to
$\theta$.  
Here, we set the time step $\Delta t = 0.0002$, the regularized
parameter $\delta = 0.15$, and the weighting factor ${\tilde \alpha} =
-0.05$. We adopt the trapezoidal rule and the forth-order Runge-Kutta
method for the spatial and temporal integration,
respectively. For detailed numerical schemes, refer to 
\citet{matsuoka06} and references therein.  

%% The reference list follows the main body and any appendices.
%% Use LaTeX's thebibliography environment to mark up your reference list.
%% Note \begin{thebibliography} is followed by an empty set of
%% curly braces.  If you forget this, LaTeX will generate the error
%% "Perhaps a missing \item?".
%%
%% thebibliography produces citations in the text using \bibitem-\cite
%% cross-referencing. Each reference is preceded by a
%% \bibitem command that defines in curly braces the KEY that corresponds
%% to the KEY in the \cite commands (see the first section above).
%% Make sure that you provide a unique KEY for every \bibitem or else the
%% paper will not LaTeX. The square brackets should contain
%% the citation text that LaTeX will insert in
%% place of the \cite commands.

%% We have used macros to produce journal name abbreviations.
%% AASTeX provides a number of these for the more frequently-cited journals.
%% See the Author Guide for a list of them.

%% Note that the style of the \bibitem labels (in []) is slightly
%% different from previous examples.  The natbib system solves a host
%% of citation expression problems, but it is necessary to clearly
%% delimit the year from the author name used in the citation.
%% See the natbib documentation for more details and options.

%\bibliography{apj-jour,reference} 
\bibliographystyle{apj}

\clearpage

\end{document}